\newif\ifpreprint
\preprinttrue 

\newif\ifSupplementary
\Supplementaryfalse
\ifpreprint
\documentclass{natureprintstyle}
\else
\documentclass{nature}
\fi
\usepackage{subfigure}
\usepackage{graphicx}
\usepackage{multirow}
\usepackage{amssymb,hyperref}
\def\msun{$M_{\odot}$} 
\def\lsun{$L_{\odot}$}

\newcommand\ion[2]{#1$\;${\small\rmfamily #2}\relax}%
\def\araa{Annu.~Rev.~Astron.~Astrophys.}
\def\aap{Astron.~Astrophys.}            
\def\aj{Astron.~J.}                     
\def\apj{Astrophys.~J.}                 
\def\apjl{Astrophys.~J.}                
\def\apjs{Astrophys.~J.~Suppl.~Ser.}    
\def\mnras{Mon.~Not.~R.~Astron.~Soc.}   
\def\nat{Nature}                        
\def\pasp{Publ.~Astron.~Soc.~Pacif.}    
\def\ssr{Space~Science~Reviews}

\def\aapr{Astron.~Astrophys.~Rev}            

\usepackage{multibib}
\newcites{maintext}{\mbox{ }}
\newcites{method}{\mbox{ }}
\newcites{supplementary}{\mbox{ }}

\newcommand\arcmin{\mbox{$^\prime$}}%
\newcommand\arcsec{\mbox{$^{\prime\prime}$}}%
\newcommand\farcs{\mbox{$.\!\!^{\prime\prime}$}}
\newcommand\farcm{\mbox{$.\!\!^{\prime}$}}

\title{A likely decade-long sustained tidal disruption event}

\author{Dacheng Lin$^{1}$,
  James Guillochon$^{2}$,
  S. Komossa$^{3}$,
  Enrico Ramirez-Ruiz$^{4}$,
  Jimmy A. Irwin$^{5,6}$,
  W. Peter Maksym$^{7}$,
  Dirk Grupe$^{8}$,
  Olivier Godet$^{9,10}$,
  Natalie A. Webb$^{9,10}$,
  Didier Barret$^{9,10}$,
  B. Ashley Zauderer$^{11}$,
  Pierre-Alain Duc$^{12}$,
  Eleazar R. Carrasco$^{13}$,
  Stephen D. J. Gwyn$^{14}$
}

\begin{document}

\maketitle

\begin{affiliations}
\item Space Science Center, University of New Hampshire, Durham, NH 03824, USA
\item Harvard-Smithsonian Center for Astrophysics, The Institute for Theory and Computation, 60 Garden Street,
Cambridge, MA 02138, USA
\item QianNan Normal University for Nationalities, Longshan Street, Duyun City of Guizhou Province, China
\item Department of Astronomy and Astrophysics, University of California, Santa Cruz, CA 95064, USA
\item Department of Physics and Astronomy, University of Alabama, Box 870324, Tuscaloosa, AL 35487, USA
\item Department of Physics and Astronomy, Seoul National University, Seoul 08826, Korea
\item Harvard-Smithsonian Center for Astrophysics, 60 Garden St., Cambridge, MA 02138, USA
\item Space Science Center, Morehead State University, 235 Martindale Drive, Morehead, KY 40351, USA
\item CNRS, IRAP, 9 avenue du Colonel Roche, BP 44346, F-31028 Toulouse Cedex 4, France
\item Universit\'{e} de Toulouse, UPS-OMP, IRAP, Toulouse, France
\item Center for Cosmology and Particle Physics, New York University,
  4 Washington Place, New York, NY 10003 USA
\item AIM Paris-Saclay Service d'astrophysique, CEA-Saclay, 91191 Gif
  sur Yvette, France
\item Gemini Observatory/AURA, Southern Operations Center, Casilla 603, La Serena, Chile
\item Canadian Astronomy Data Centre, Herzberg Institute of Astrophysics, 5071 West Saanich Road, Victoria, British Columbia, V9E 2E7, Canada
\end{affiliations}

\begin{abstract}
  Multiwavelength flares from tidal disruption and accretion of stars
  can be used to find and study otherwise dormant massive black holes
  in galactic nuclei\cite{re1988}. Previous well-monitored candidate
  flares are short-lived, with most emission confined to within
  $\sim$1 year\cite{gechre2012,zabema2013,mikami2015,vaanst2016}. Here
  we report the discovery of a well observed super-long ($>$11 years)
  luminous soft X-ray flare from the nuclear region of a dwarf
  starburst galaxy. After an apparently fast rise within $\sim$4
  months a decade ago, the X-ray luminosity, though showing a weak
  trend of decay, has been persistently high at around the Eddington
  limit (when the radiation pressure balances the gravitational
  force). The X-ray spectra are generally soft (steeply declining
  towards higher energies) and can be described with Comptonized
  emission from an optically thick low-temperature corona, a
  super-Eddington accretion signature often observed in accreting
  stellar-mass black holes\cite{mimima2013}. Dramatic
  spectral softening was also caught in one recent observation,
  implying either a temporary transition from the super-Eddington
  accretion state to the standard thermal state or the presence of a
  transient highly blueshifted ($\mathbf{\sim0.36c}$) warm
  absorber. All these properties in concert suggest a tidal disruption
  event (TDE) of an unusually long super-Eddington accretion phase
  that has never been observed before.
\end{abstract}

The X-ray source 3XMM~J150052.0+015452 (XJ1500+0154 hereafter) was
serendipitously detected in frequent observations\cite{ranujo2015} of
the foreground galaxy group NGC 5813 by the X-ray observatories \emph{Chandra}
and \emph{XMM-Newton} from 2005 to 2011. Our follow-up observation of
the source with \emph{Chandra} on February 23rd 2015 provided a well
constrained X-ray position coincident with the center of the galaxy
SDSS J150052.07+015453.8, to within 0\farcs18 (Figure 1, see
\emph{SI}). The galaxy lies at a redshift of 0.145, or a luminosity
distance of $D_L=689$ Mpc (for $H_0$=70 km~s$^{-1}$~Mpc$^{-1}$,
$\Omega_\mathrm{M}=0.3$, $\Omega_{\Lambda} = 0.7$), with strong
emission lines indicative of intense star-forming activity. It has a
total stellar mass of $\sim$$6\times10^9$ \msun\ (see \textit{SI}),
comparable to that of the Large Magellanic Cloud. For such a small
galaxy, we expect\cite{revo2015} the central supermassive black hole
(SMBH) to have mass $\sim$$10^6$ \msun.

The upper panel in Figure~2 shows the long-term evolution of the X-ray
luminosity $L_{\rm X}$. Our best fits to the spectra with sufficient
counts are shown in the lower panel of the figure and are given in
Table~\ref{tbl:spfit}. One striking property of the source is the
fast-rise-very-slow-decay outburst profile. It was not detected in the
first \emph{Chandra} observation on April 2nd 2005, with $L_{\rm
  X}<4.3\times10^{41}$ erg s$^{-1}$ (3$\sigma$ upper limit, assuming a
powerlaw source spectrum of photon index 2.0). Less than 4 months
later (July 23rd 2005), the source was detected in the first
\emph{XMM-Newton} observation, with $L_\mathrm{X}\sim5.5\times10^{42}$
erg s$^{-1}$. It was detected at an even higher luminosity three years
later, in one \emph{Chandra} observation on June 5th 2008 and two
\emph{XMM-Newton} observations in February 2009, with $L_{\rm
  X}\sim7.0\times10^{43}$ erg s$^{-1}$. The luminosity decreased only
slightly to $\sim3.0\times10^{43}$ erg s$^{-1}$ in seven
\emph{Chandra} observations in March--April 2011. Similar luminosities
were seen in our follow-up observations later, one by \emph{Swift} on
March 28th 2014, one by \emph{Chandra} on February 23rd 2015, and
seven by \emph{Swift} in February 2016. Given the probably small
central SMBH and correcting for emission outside the X-ray band, the
source luminosity has been most likely at around the Eddington limit
since it went into the outburst.

Another special property of the source is the generally quasi-soft
X-ray spectra, most evident in the \textit{XMM-Newton} and
\textit{Chandra} observations in 2008--2011. These spectra can be
roughly described with a dominant thermal disk of apparent maximum
disk temperature $kT_\mathrm{diskbb}\sim0.3$ keV plus a weak powerlaw
(see \textit{SI}). However, such a model is physically unacceptable,
because the standard thin accretion disk around a SMBH is expected to
produce much cooler thermal emission ($kT_\mathrm{diskbb}\lesssim0.1$
keV)\cite{gido2004}. Instead, the spectra can be described well with
the Comptonization model CompTT\cite{ti1994}, with an optically thick
($\tau\sim 4$--11) low-temperature ($kT_\mathrm{e}\sim0.4$--1.3 keV)
corona (see \textit{SI}). Such spectral parameters are commonly seen
in ultraluminous X-ray sources
(ULXs)\cite{glrodo2009,liirwe2013,mimima2013}, most of
which are believed to be super-Eddington accreting stellar-mass black
holes, except that XJ1500+0154 had orders of magnitude higher
luminosities. Therefore, we identified the observations in 2008--2011
as being in the super-Eddington accretion state.

Surprisingly, we obtained a much softer X-ray spectrum in the
\textit{Chandra} observation in 2015, mostly due to a drop (by a
factor of 7.0) in the count rate above $\sim$1 keV (Figure 2),
compared with the observations in 2008--2011. This super-soft spectrum
can be described with a dominant thermal disk of
$kT_\mathrm{diskbb}\sim0.13\pm0.01$ keV plus a very weak powerlaw
(Figure 2). Such a cool thermal disk is expected from accretion onto a
SMBH below the Eddington limit. Then the source could be in the
thermal state, which in turn supports the identification of the
super-Eddington accretion state in previous observations. In this case
the X-ray spectral evolution of XJ1500+0154 is very similar to a
transient ULX in M31\cite{mimima2013}, which changed from a
super-Eddington accretion state to a thermal state within 20 days with
the X-ray luminosity decreasing only slightly. However, we find that
the spectrum can be described equally well with the CompTT model that
fits the \textit{Chandra} observations in 2011, except for additional
absorption by a strong ($N_\mathrm{H}\sim6\times10^{23}$ cm$^{-2}$)
ionized ($\log(\xi)=2.8$) absorber with a blueshifted velocity of
$0.36c$. Powerful sub-relativistic winds are expected in
super-Eddington accreting black holes\cite{kimu2016}, and highly
blueshifted warm absorbers have been detected in
ULXs\cite{pimifa2016}. Therefore, this interpretation for the
\textit{Chandra} observation in 2015 also supports the identification
of the super-Eddington accretion state in previous observations. The
recent \emph{Swift} observations in 2016 have poor statistics, but
some of them did not seem to show similar super-soft X-ray
spectra. Therefore, the source has not completely settled to a new
state of super-soft X-ray spectra, and the super-Eddington accretion
seems to have lasted for $\gtrsim$$11$ years (see \textit{SI}).

No sign of persistent nuclear activity is seen in the optical emission
lines of the host galaxy, whose ratios are fully consistent with those
expected from a starburst galaxy. There are many other properties of
the source that argue against the possibility that it is a standard
active galactic nucleus (AGN, see \textit{SI}). In particular, no AGN
is known to show X-ray spectra as soft as XJ1500+0154 within the
1--4.5 keV energy band or show dramatic quasi-soft to super-soft X-ray
spectral change. The large X-ray variability (a factor of $>$97) is
also extremely rare among AGNs. Therefore, although we cannot
completely rule out that XJ1500+0154 is just a highly variable AGN at
this point, its X-ray outburst is best explained as tidal disruption
of a star by the central black hole. This interpretation is strongly
supported by our new discovery of two other sources that seemed to be
in X-ray outbursts with similar quasi-soft X-ray spectra as
XJ1500+0154 but have host galaxies showing no sign of nuclear activity
in optical (see \textit{SI}).

The super-Eddington accretion phase from tidal disruption of a
solar-type star by a $10^6$ \msun\ black hole can last
$t_\mathrm{Edd}\approx 2$ years, with the peak mass accretion rate
highly super-Eddington\cite{re1988,ul1999}.  One main property of a
super-Eddington accretion disk is a lower radiative efficiency than a
standard thermal thin disk, due to significant super-Eddington effects
of photon trapping and outflows in the inner disk
region\cite{ohmi2007,krpi2012,kimu2016}. These effects are more
serious at higher accretion rates, with the disk luminosity sustained
at around the Eddington limit. The Eddington-limited slow decay of our
source thus agrees well with the super-Eddington accretion signatures
suggested by the X-ray spectra. The long super-Eddington accretion
phase of $\gtrsim$$11$ years in our event would imply disruption of a
very massive star (10 \msun) based on the standard
theory\cite{ul1999}. However, it has been realized that the evolution
of TDEs heavily depends on how the streams of tidal debris intersect
each other\cite{ko1994,gura2015,pisvkr2015,shkrch2015,hastlo2016}.  It
should be common for circularization of the fallback mass onto the
accretion disk to occur at a much larger distance, resulting in a much
longer viscous time scale, than predicted from the standard
theory\cite{gura2015}. Therefore, $t_\mathrm{Edd}$ can be very long in
a slow circularization process, unless the circularization is so slow
that the peak accretion rate drops to be sub-Eddington.

We plot in Figure~2 (solid line) the evolution of the luminosity from
a full disruption of a 2 \msun\ star by a $10^6$ \msun\ black hole,
with the accretion of the mass slowed relative to the fallback time by
3 years. The super-Eddington effects were taken into account by
introducing a logarithmic dependence of the radiative efficiency on
the accretion rate above the Eddington limit\cite{kimu2016} (see
\textit{SI}). We assumed that 25\% of the radiation is in X-rays, as
inferred from the spectral modeling. The model describes the data
well. The total energy release and the total mass accreted onto the
black hole until the last \textit{Swift} observation would then be
$6.4\times10^{52}$ ergs and 0.89 \msun, respectively, which are orders
of magnitude higher than seen in other known
events\cite{liname2002,kohasc2004,vaanst2016}.

Although disruption of a very massive star of 10 \msun\ with prompt
circularization can also describe the data, such disruption is
expected to be orders of magnitude rarer than disrupting a star
of 2 \msun\ with slow circularization (see \textit{SI}). Therefore our
event provides the first convincing evidence of slow circularization
effects in TDEs, which are expected to be very common when the black
holes are small ($\sim$$10^6$ \msun) but were not clearly observed
before probably due to observational bias\cite{gura2015}.

We calculated the rate of events similar to XJ1500+0154 to be
$\sim$$4\times10^{-7}$ per galaxy per year (see \textit{SI}), which is
about two orders of magnitude lower than estimated for short
TDEs\cite{dobrer2002}. One main reason for the low rate of events like
XJ1500+0154 could be the relatively large mass (2\msun) of the
disrupted stars required, which is only possible in starburst
galaxies\cite{ko2016}.  Although events like XJ1500+0154 are rare,
their extreme duration and radiative inefficiency mean that their
contributions to the luminosity function of active galactic nuclei and
to the growth of the black holes are comparable to or even higher than
those of short events. TDEs with a shorter super-Eddington accretion
phase than XJ1500+0154 could be more common. The discovery of our
event opens up a new realm in which to search for super-Eddington
accreting TDEs, that is, by investigating sources with quasi-soft
X-ray spectra. Our discovery of the other two candidates is the result
of applying this scheme.

This is the first time that X-ray spectra resembling typical
super-Eddington accreting stellar-mass black holes were observed in an
accreting SMBH. If our interpretation of a super-Eddington accreting
TDE for XJ1500+0154 is correct, it would have important implications
for the growth of massive black holes. The detection of quasars at
redshift $z>6$ with black hole masses $\sim10^9$ \msun\ poses a
problem to explain their growth with accretion via a standard thin
disk at the Eddington rate\cite{mowave2011}. However, their formation
would be possible if the black holes can accrete at a super-Eddington
rate during an early phase\cite{vore2005}. Our event shows that
super-Eddington accretion onto massive black holes can occur, giving
strong observational support to this model. The high absorption in
these systems would mean that the search for them should be through
radio and infrared\cite{marala2005}, because their X-ray spectra, if
as soft as XJ1500+0154, would be completely absorbed.

We expect the accretion rate to drop by an order of magnitude to be
well sub-Eddington in the next ten years based on our model of full
disruption of a 2 \msun\ star. By continued monitoring of the event,
we will be able to test our TDE interpretation and to determine the
duration of the super-Eddington accretion phase and the origin of
spectral softening. We will also be given a rare opportunity to
observe the spectral evolution of the event across different accretion
regimes and to investigate its connection with short super-soft events
that are mostly believed to accrete below the Eddington limit.

\begin{addendum}

\item[Acknowledgments] D.L. is supported by the National Aeronautics
  and Space Administration through Chandra Award Number GO5-16087X
  issued by the Chandra X-ray Observatory Center, which is operated by
  the Smithsonian Astrophysical Observatory for and on behalf of the
  National Aeronautics Space Administration under contract
  NAS8-03060. We thank the \textit{Swift} PI Neil Gehrels for
  approving our ToO request to make several observations of XJ1500+0154.
  
\item[Author Contributions]

  D.L.  wrote the main manuscript and led the data analysis.
  J. G. helped with the modeling of the long-term X-ray light curve
  and wrote the text on the modeling in the Supplementary
  Information. S. G. stacked the CFHT images. All authors discussed
  the results and commented on the manuscript.

\item[Competing Interests] 
The authors declare that they have no competing financial interests.

\item[Author Information] 
Correspondence and requests for materials should be addressed to
D.L. (dacheng.lin@unh.edu). 
\end{addendum}

\clearpage

\begin{table*}
\centering
\caption{\textbf{Spectral fit results for high-quality spectra.}  The
  X1 and C10 spectra were rebinned to have at least one count per bin
  and were fitted by minimizing the C statistic, while the X2, X3, C2,
  and C3--C9 spectra were rebinned to have at least 20 counts per bin
  and were fitted by minimizing the $\chi^2$ statistic. The fits used
  energy channels within 0.3--10 keV for \textit{XMM-Newton} and energy channels
  within 0.3--7 keV for \textit{Chandra}.  All models include Galactic
  absorption of column density $N_\mathrm{H, Gal}=4.4\times10^{20}$
  cm$^{-2}$. The absorption intrinsic to the X-ray source at redshift
  0.14542 was also included and fixed at $N_\mathrm{H,
    i}=4.2\times10^{21}$ cm$^{-2}$, which was the best-fitting value
  from the simultaneous fit to the C2, X3, and C3--C9 spectra. $L_{\rm
    abs}$ is the source rest-frame 0.34--11.5 keV luminosity,
  corrected for the Galactic absorption but not intrinsic absorption,
  and $L_{\rm unabs}$ is the source rest-frame 0.34--11.5 keV
  luminosity, corrected for both Galactic and intrinsic
  absorption. Both $L_{\rm abs}$ and $L_{\rm unabs}$ are in units of
  10$^{43}$ erg s$^{-1}$. All errors are at the 90\%-confidence
  level. Parameters without errors were fixed in the fits. For C10, two
  models were tested: diskbb+PL and zxipcf(comptt). For the latter
  model, the CompTT component was fixed at the best fit of this model
  to C3--C9, and the luminosities $L_{\rm abs}$ and $L_{\rm unabs}$
  was simply copied from those of C3--C9. The reduced $\chi^2$ values
  are given for fits using the $\chi^2$ statistic, but not for those using
  the C statistic. }
\label{tbl:spfit}
\bigskip
\scriptsize
\sffamily
\begin{tabular}{r|c|c|c|cc}
\hline
\hline
Obs. & Model& Parameters  & $\chi^2_\nu(\nu)$& $L_{\rm abs}$ & $L_{\rm unabs}$\\

\hline
\multirow{2}{*}{X1} &\multirow{2}{*}{diskbb+PL}  & {$kT_{\rm diskbb}=0.10^{+ 0.06}_{-0.04}$ keV, $N_\mathrm{diskbb}=210^{+27841}_{-204}$, }&\multirow{2}{*}{...}& \multirow{2}{*}{$ 0.08^{+ 0.04}_{-0.03}$}  & \multirow{2}{*}{$ 0.55^{+ 0.95}_{-0.31}$} \\
&& $\Gamma_\mathrm{PL}=2.5$, $N_{\rm PL}=4.6^{+ 3.9}_{-3.4}\times10^{-6}$ & \\
X2 & CompTT & {$kT_\mathrm{0}=0.04$ keV, $kT_\mathrm{e}=0.35^{+ 0.08}_{-0.05}$ keV,  $\tau=10.8^{+  2.4}_{ -2.1}$} &$1.09(106)$ & $ 1.16^{+ 0.06}_{-0.06}$  & $ 6.18^{+ 0.65}_{-0.60}$ \\
X3 & CompTT & {$kT_\mathrm{0}=0.04$ keV, $kT_\mathrm{e}=0.51^{+ 0.30}_{-0.13}$ keV,  $\tau=7.3^{+  2.1}_{ -2.2}$} &$0.90( 99)$ & $ 1.13^{+ 0.06}_{-0.06}$  & $ 7.03^{+ 0.71}_{-0.66}$ \\
C2 & CompTT & {$kT_\mathrm{0}=0.04$ keV, $kT_\mathrm{e}=0.51^{+ 0.33}_{-0.14}$  keV,  $\tau=7.6^{+  3.0}_{ -2.5}$} &$0.90( 43)$ & $ 1.08^{+ 0.09}_{-0.09}$  & $ 6.24^{+ 1.55}_{-1.32}$ \\
C3--C9 & CompTT & {$kT_\mathrm{0}=0.04$ keV, $kT_\mathrm{e}=1.33^{+ 1.48}_{-0.43}$ keV,  $\tau=3.9^{+  1.3}_{ -1.7}$} &$0.83(138)$ & $ 0.60^{+ 0.02}_{-0.02}$  & $ 3.38^{+ 0.23}_{-0.21}$ \\
\multirow{4}{*}{C10} & \multirow{2}{*}{diskbb+PL} & {$kT_{\rm diskbb}=0.13^{+ 0.01}_{-0.01}$ keV, $N_\mathrm{diskbb}=312^{+310}_{-154}$,} & \multirow{2}{*}{...}& \multirow{2}{*}{$ 0.31^{+ 0.05}_{-0.04}$}  & \multirow{2}{*}{$ 2.88^{+ 0.86}_{-0.66}$} \\
&&  $\Gamma_\mathrm{PL}=2.5$, $N_{\rm PL}=2.2^{+ 1.7}_{-1.2}\times10^{-6}$ &&\\
& \multirow{2}{*}{zxipcf(CompTT)} & $kT_\mathrm{0}=0.04$ keV, $kT_\mathrm{e}=1.33$ keV,  $\tau=3.9$, & \multirow{2}{*}{...}& \multirow{2}{*}{$ 0.60^{+ 0.02}_{-0.02}$}  & \multirow{2}{*}{$ 3.38^{+ 0.23}_{-0.21}$} \\
&&$N_\mathrm{H}=64\pm10\times10^{22}$ cm$^{-2}$, $\log(\xi)=2.78\pm0.04$, $z=-0.36\pm0.02$ &\\
\hline
\hline
\end{tabular}
\end{table*}

\clearpage

\begin{figure*}
\begin{center}
\includegraphics[width=3.5in]{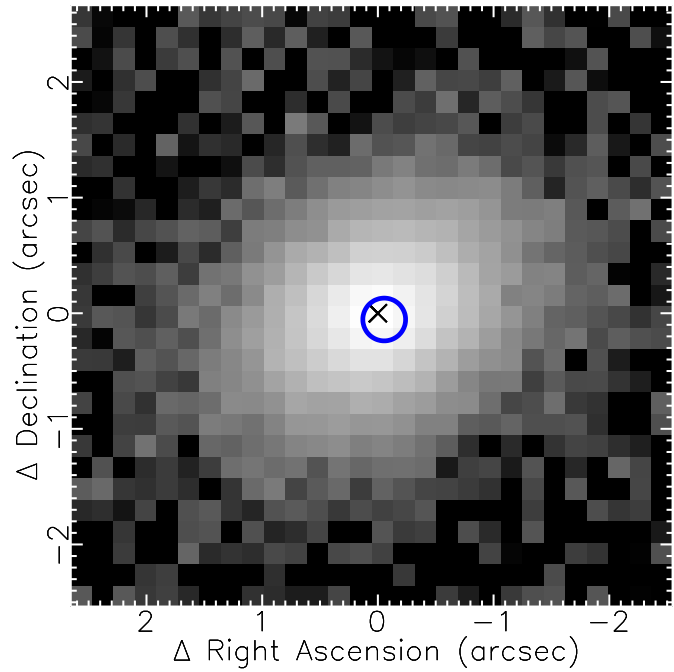}
\end{center}
\vskip -0.2in
\caption{
  \textbf{The CFHT/MegaPrime
  $r\arcmin$-band image around the field of XJ1500+0154 indicates its
  galactic nuclear origin.} The origin of the image is at the center
of the galaxy SDSS J150052.07+015453.8 (black cross). The blue circle
of radius 0\farcs18 (0.5 kpc) represents the 95\% positional
uncertainty of XJ1500+0154. \label{fig:cfhtimg}}
\end{figure*}

\clearpage

\begin{figure*}
\begin{center}
\includegraphics[width=5.8in]{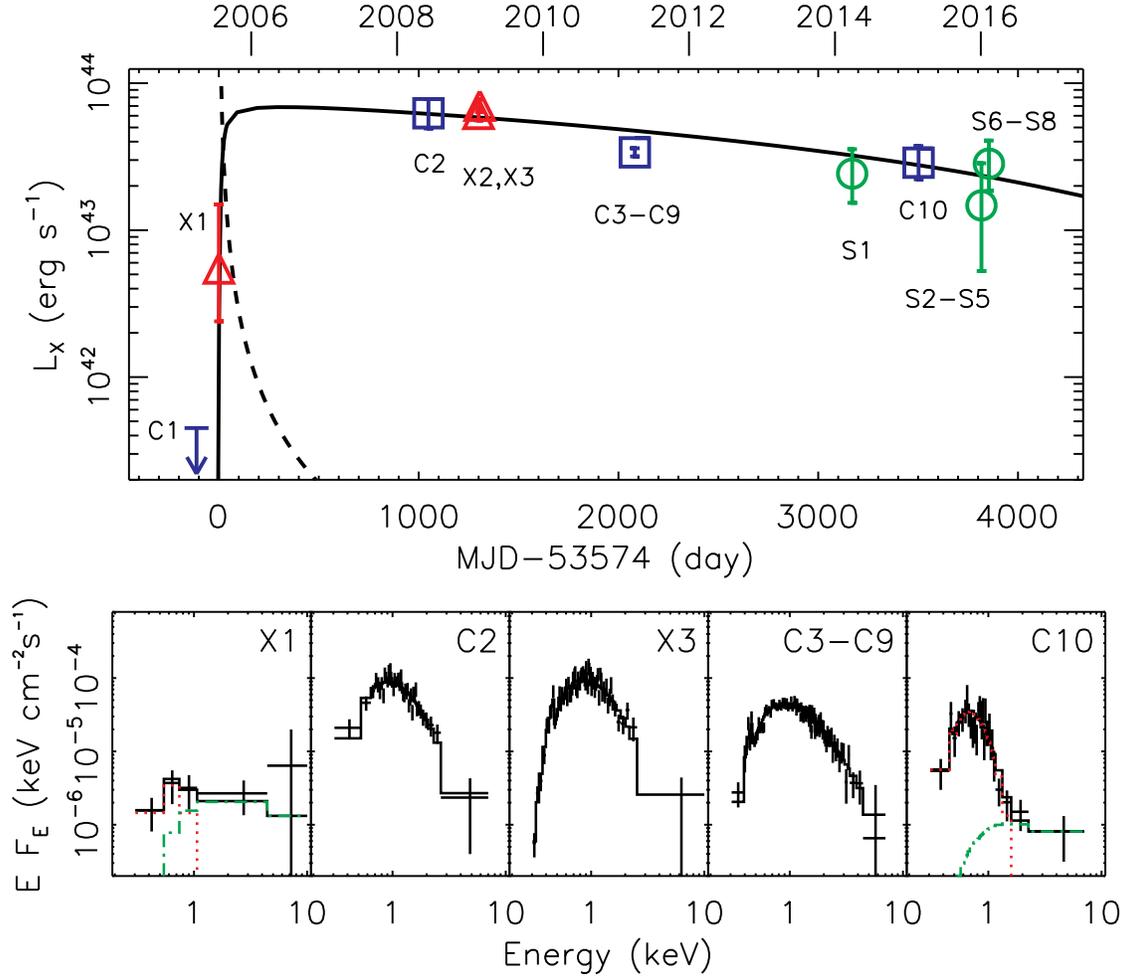}
\end{center}
\vskip -0.2in
\caption{\textbf{The long-term evolution of the X-ray luminosity and
    spectrum of XJ1500+0154.} Upper panel: The long-term source
  rest-frame 0.34--11.5 keV unabsorbed luminosity curve. The
  \emph{Chandra}, \emph{XMM-Newton} and \emph{Swift} observations are
  shown as blue squares, red triangles and green circles,
  respectively, with 90\% error bars, but for the first
  \textit{Chandra} observation C1 in 2005 the $3\sigma$ upper limit is
  shown with an arrow. We have merged the seven \emph{Chandra}
  observations in 2011 to create a single coadded spectrum, given the
  lack of significant spectral/flux change in these
  observations. Similarly we also created a coadded spectrum from the
  combination of S2--S5 and another one from S6--S8. For clarity, we
  have offset S2--S5 to be one month earlier, because they are too
  close to S6--S8 in time.  The solid line is a model of disrupting a
  2 \msun\ star by a black hole of mass $10^6$ \msun\ with slow
  circularization and super-Eddington effects (see \textit{SI}). Such a
  model describes the data well.  The dashed line plots $t^{-5/3}$,
  assuming a peak X-ray luminosity of $10^{44}$ erg s$^{-1}$ that is
  reached two months after disruption of the star; it represents the
  typical evolution trend for thermal
  TDEs\cite{kohasc2004,mauler2013}, which obviously last much shorter
  than our event.  Lower panels: The unfolded X-ray spectra. The X1
  and C10 observations were fitted with a diskbb model (red dotted
  line) plus a PL (green dot-dashed line), and the C2, X2, X3, and
  C3--C9 observations were fitted with a CompTT model (the X2 spectrum
  is not shown but looks very similar to X3). Note that C10
  can also be described with the CompTT fit to C3--C9 subject to a
  fast warm absorber. For clarity, we show only pn data for the
  \emph{XMM-Newton} observations. Also for clarity, the spectra were
  rebinned to be above 2$\sigma$ in each bin in the
  plot. \label{fig:lumlcsp}}
\end{figure*}

\clearpage

\Supplementarytrue

\section*{Method}

\noindent \textbf{XMM-Newton observations and data analysis}

XJ1500+0154 was serendipitously detected at off-axis angles of
$\sim$13$\arcmin$ in three \emph{XMM-Newton} observations (X1--X3
hereafter; see Supplementary Table~1) of NGC 5813. X1 was made in
July 2005, while X2 and X3 were made in February 2009, only six days
apart. The source was detected in all three European Photon
Imaging Cameras (i.e., pn, MOS1/M1, and
MOS2/M2)\cite{jalual2001,stbrde2001,tuabar2001} in the imaging mode in
X1, but only in pn and MOS2 in X2 and X3 because the source is outside
the field of view (FOV) of MOS1 in these two observations. We used SAS
15.0.0 and the calibration files of 2016 February for reprocessing the
X-ray event files and follow-up analysis. The data in strong
background flare intervals, seen in all cameras in all observations,
were excluded following the SAS thread for the filtering against high
backgrounds. The final clean exposures used are given in
Supplementary Table~1. We extracted the source spectra from all
available cameras using a circular region of radius 20$\arcsec$. The
background spectra were extracted from a large circular region near
the source, using a radius of 100$\arcsec$ for MOS1 and MOS2 and a
radius of 50$\arcsec$ for pn. The event selection criteria followed
the default values in the pipeline\cite{wascfy2009}. For X2 and X3, in
which the source was bright, we also extracted the pn light curves
binned at 500 s using the SAS tool \texttt{epiclccorr} and performed
variability tests using the \texttt{ekstest} tool. We used the 0.3--3
keV band, where the source counts dominated over those of the
background. \\

\noindent \textbf{Chandra observations and data analysis}

XJ1500+0154 was serendipitously covered in nine \emph{Chandra}
observations (C1--C9 hereafter; see Supplementary Table~1) of NGC
5813, but all at large pointing offsets
(11$\arcmin$--15$\arcmin$). The dense observations in 2011 (C3-C9,
$\sim$0.5 Ms) are from a Large Program (LP, PI: Dr. Scott Randall) on
NGC 5813. All nine observations used the imaging array of the
AXAF CCD Imaging Spectrometer (ACIS)\cite{bapiba1998}, and XJ1500+0154
fell in the back-illuminated chip S1 in all observations except C2, in
which it fell in the front-illuminated chip I3.  We had a
\textit{Chandra} follow-up observation of the source for 37 ks in February 2015 (C10 hereafter), with the aim point at the back-illuminated
chip S3.  We reprocessed all the data to apply the latest calibration
(CALDB 4.6.7) using the script \textit{chandra\_repro} in the
\textit{Chandra} Interactive Analysis of Observations (CIAO, version
4.7) package. No clear background flares were seen, and we used all
data for all observations.

The spectra of XJ1500+0154 were extracted for each observation. We used
a circular source region enclosing 90\% of the point spread function
(PSF) at 1.0 keV and a circular background region of a radius of
$50\arcsec$ near the source. However, there are three exceptions. For
observations C1 and C2, we used the 70\% PSF radius for the source
region, considering that the source was not detected in C1 and was
near the CCD edge in C2. For our follow-up observation C10, in which
the source is near the aim point with minimum background
contamination, we used the 95\% PSF radius for the source region.  We
used the CIAO task \texttt{mkacisrmf} to create the response matrix
files and the CIAO tasks \texttt{mkarf} and \texttt{arfcorr} to create
the point-source aperture corrected auxiliary response
files. Considering no significant difference between LP observations,
which were taken within 13 days, and in order to improve the
statistics for spectral modeling, we created a single spectrum
combining LP observations.  For observation C1 in which the source was
not detected, we used the CIAO task \texttt{aprates} to determine
confidence bounds of the flux.

We measured the short-term variability within 0.4--3 keV adopting the
Gregory-Loredo algorithm\cite{grlo1992} implemented by the CIAO tool
\texttt{glvary}\cite{evprgl2010}. It splits the events into multiple
time bins and looks for significant deviations. The variation of the
effective area with time was taken into account and was obtained by
another CIAO tool \texttt{dither\_region}. The different degrees of
confidence is indicated by the parameter of ``variability index'',
which spans values within [0, 10] and is larger for variability of
higher confidence\cite{evprgl2010}.

Our C10 observation was intended to provide an accurate position of
XJ1500+0154, utilizing the sub-arcsec resolution of \textit{Chandra}
near the aimpoint. We performed the source detection by applying the
CIAO \texttt{wavdetect} wavelet-based source detection
algorithm\cite{frkaro2002} on the 0.3--7 keV image binned at single
sky pixel resolution. We then carried out the absolute astrometric
correction for the X-ray sources by cross-correlating them with
optical sources in the Canada-France-Hawaii Telescope (CFHT)
MegaPrime/MegaCam\cite{bochab2003} $r\arcmin$-band stacked images. We
only used 19 matches that are outside the strong diffuse gas emission
in NGC 5813\cite{rafogi2011} and have X-ray 95\% statistical
positional errors $\le$1\farcs2 (based on Equation 12 in Kim et
al.\cite{kikiwi2007}) and magnitude $m_{r\arcmin}<24.0$ AB mag for
astrometric correction.  These 19 matches do not include XJ1500+0154,
in order to reduce the effect of the astrometric correction on the
identification of its optical counterpart. The astrometric correction
method in Lin et al.\cite{licawe2016} was used, by searching for the
translation and rotation of the X-ray frame that minimize the total
$\chi^2$ ($\chi$ is the ratio of the X-ray-optical separation to the
total positional error) for 90\% (i.e., 17, allowing 2 matches to be
spurious or bad) of the matches that have the smallest $\chi$
values. The uncertainties of the translation and rotation and thus the
systematic positional errors of the X-ray sources were estimated using
200 simulations. In order to calculate the statistical positional
uncertainty for the source, we carried out 2000 ray-trace simulations
with \texttt{MARX} 5.1.0 at positions near it and at the same off-axis
angle. The spectrum from the thermal disk plus powerlaw fit to C10 was
assumed.\\

\noindent \textbf{Swift observation and data analysis}

At our request, \textit{Swift}\cite{gechgi2004} observed XJ1500+0154
in two epochs: one observation on March 28th 2014 (S1 hereafter), and
seven observations between February 3rd--14th 2016 (S2--S8 hereafter,
Supplementary Table~1). We analyzed the data with FTOOLS 6.18 and the
calibration files released on July 31st 2015.  The X-ray telescope
(XRT)\cite{buhino2005} was operated in the Photon Counting mode for
all observations, and we reprocessed the data with the task
\texttt{xrtpipeline} (version 0.13.2). The spectrum was extracted,
using radii of 25$\arcsec$ and $2\arcmin$ for the circular source and
background regions, respectively. The source was hardly detected by
the XRT in S2--S5, and we created a co-added XRT spectrum from these
observations. The source was clearly detected in S6--S8, and we also
created a co-added XRT spectrum from these observations. The source
net count rate was higher in S6--S8 than in S2--S5 at the $3\sigma$
confidence level.

The UV-Optical Telescope (UVOT)\cite{rokema2005} in S1 used three
standard UV filters W1 (2.3 ks), M2 (2.3 ks), and W2 (2.3 ks). In
S2--S8 the UVOT used the ``Filter of the Day'' mode, and we combined
images for different filters, resulting in total exposures of 6.3 ks,
1.4 ks, 3.9 ks and 3.8 ks for the U, W1, M2, and W2 filters,
respectively. To obtain the photometry, we used the task
\textit{uvotsource} with radii of 5$\arcsec$ and 20$\arcsec$ for the
circular source and background regions, respectively.\\

\noindent \textbf{ROSAT Observations}

XJ1500+0154 was in the field of view of a \emph{ROSAT} High-Resolution
Imager pointed observation in 1998 for 1.9 ks
(Supplementary Table~1). XJ1500+0154 was not detected in the observation,
and we extracted the source and background spectra using a circular
region of radius 10$\arcsec$ and a circular region of radius
$50\arcsec$, respectively, in order to estimate the flux limit.\\

\noindent \textbf{CFHT MegaCam observations and data analysis}

There are seven CFHT/MegaPrime $r\arcmin$-band images, with exposure
345 second each and taken on May 5th 2014, and eight $g\arcmin$-band
images, with exposure 345 second each and taken on April 26th 2014. We
produced two stacked images, one for the $r\arcmin$ band (the final
seeing FWHM is 0\farcs65) and the other for the $g\arcmin$ band (the
final seeing FWHM is 0\farcs80), using MegaPipe\cite{gw2008}, and
aligned their astrometry to the Sloan Digital Sky Survey
(SDSS)\cite{abadag2009}. The sources detected from these two stacked
images were used to compare with the SDSS photometry to detect the
source optical variability and to compare with X-ray sources to align
the X-ray source astrometry. We also fitted the host galaxy profile in
the CFHT/MegaPrime stacked images using GALFIT\cite{pehoim2010}. Ten
stars within $2\arcmin$ from XJ1500+0154 were used to construct the
point spread function of the images. \\

\noindent \textbf{X-ray spectral fits}

The X-ray activity of XJ1500+0154 should be due to accretion onto a
BH, and we fitted the X-ray spectra with several models typically used
to study such an object. Given that XJ1500+0154 is most likely
associated with a galaxy at $z=0.14542$, we applied this redshift to
all the spectral models that we tested with the convolution model
\textit{zashift} in XSPEC\cite{ar1996}. All models included the
Galactic absorption\cite{kabuha2005} fixed at
$N_\mathrm{H}=4.4\times10^{20}$ cm$^{-2}$ using the \textsc{tbabs}
model. Possible absorption intrinsic to the source was accounted for
using the \texttt{ztbabs} model. The abundance tables from Wilms et
al.\cite{wialmc2000} were used. \\

\noindent \textbf{Data availability statement}
The data that support the plots within this paper and other findings of this study are available from the
corresponding author upon reasonable request.

\clearpage
\ifpreprint
\else
\setcounter{page}{1}
\fi
\setcounter{figure}{0}
\setcounter{table}{0}
\renewcommand{\thefigure}{\arabic{figure}}
\renewcommand{\thetable}{\arabic{table}}

\Supplementarytrue

\section*{Supplementary Information}

\mbox{ }

\vskip 0.05in 
\noindent

\noindent \textbf{The X-ray source position and the host galaxy}

We obtained the position of XJ1500+0154 from C10 to be
R.A.=15:0:52.068 and Decl.=+1:54:53.79, with the 95\% positional error
of 0\farcs18 (including both the statistical component and the
systematic component from the astrometric correction procedure). This
position is only 0\farcs07 away from the center of the galaxy
SDSS~J150052.07+015453.8, consistent within the uncertainty
(Figure~1).  The number density of the galaxies that
are as bright as or brighter than SDSS~J150052.07+015453.8 in the
$r\arcmin$ band within $10\arcmin$ is 0.00016 per square arcsec,
implying that the chance probability for our X-ray source to be within
$0\farcs18$ from the center of SDSS~J150052.07+015453.8 is only
0.000017. Therefore we can securely identify this galaxy as the host
of our X-ray source, with the X-ray emission most likely from the
galaxy center (the projected offset $\sim$0\farcs18 or 0.5 kpc at the
redshift of $z=0.14542$ of the galaxy; see below).

The host galaxy in the CFHT/MegaPrime $r\arcmin$ and $g\arcmin$-band
images can be fitted with a S\'{e}rsic profile, with effective radii
of 0\farcs48$\pm$0\farcs01 (1.23$\pm$0.03 kpc) and
0\farcs41$\pm$0\farcs0.01 (1.04$\pm$0.01 kpc), axis ratios of
$0.56$$\pm$$0.02$ and $0.56$$\pm$$0.01$, and indices of
$2.2$$\pm$$0.2$ and $3.0$$\pm$$0.2$, respectively. Therefore it is a
small galaxy.

The SDSS took a spectrum of the galaxy on March 3rd 2011, which is shown
in Supplementary Figure~\ref{fig:s132740gsp}. It exhibits strong narrow emission
lines at the redshift of $z=0.14542\pm0.00001$ ($D_L=689$ Mpc,
assuming a flat universe with $H_0$=70 km~s$^{-1}$~Mpc$^{-1}$ and
$\Omega_\mathrm{M}$=0.3). We fitted the spectrum with Penalized Pixel
Fitting ({\texttt{pPXF}) software\cite{caem2004}. As the stellar
  templates, we adopted the single-population models by Maraston and
  Str{\"o}mb{\"a}ck\cite{mast2011} that were created based on the
  ELODIE library of empirical stellar spectra of solar
  metallicity\cite{prso2001} and have UV extension (below around
  3950\AA) based on theoretical models\cite{mastth2009}. We chose
  these models because of their extended age coverage (54 ages from 3
  Myr to 12 Gyr), broad wavelength range (1000.2--6800.0 \AA), and
  high resolution (FWHM $\sim$ 0.55\AA). We interpolated the models to
  make them distribute uniformly in logarithm of the age from 3 Myr to
  12 Gyr with 54 grids. We fitted the SDSS spectrum over source
  rest-frame 3200-6800 \AA, considering that the spectrum at shorter
  wavelengths is very noisy. Fourteen Gaussian functions were also
  included in the fit to model the apparent gas emission lines. No
  additive or multiplicative polynomials were included in the fit. The
  spectrum was corrected for the Galactic dust
  reddening\cite{scfida1998} of $\mathrm{E(B-V)}_\mathrm{G}=0.05$ mag
  before the fit. The intrinsic reddening was left as a free fitting
  parameter and was inferred to be $\mathrm{E(B-V)}_\mathrm{i}=0.068$
  mag.

The pPXF fit inferred intrinsic width of the emission lines to be about
$\sigma=52$ km s$^{-1}$. The stellar kinematics could not be
constrained well, as expected considering the lack of strong
absorption features. Therefore in the final fit we assumed the stellar
velocity dispersion to be the same as that of the gas. The best-fit
model is shown in Supplementary Figure~\ref{fig:s132740gsp}. The gas emission lines
are all generally consistent with a single Gaussian profile. The
intrinsic reddening of the emission lines is
$\mathrm{E(B-V)}_\mathrm{i}=0.41$ mag, assuming an intrinsic
H$\alpha$/H$\beta$ ratio of 2.85. Based on more careful Gaussian fits
to the star light subtracted spectrum and adopting this amount of
intrinsic reddening, we obtained the extinction-corrected line ratios
of $\log([{\rm O III}]\lambda5007/{\rm H}\beta)=0.18\pm0.02$
($1\sigma$ error), $\log([{\rm N II}] \lambda6583/{\rm
  H}\alpha)=-0.63\pm0.02$, $\log([{\rm S II}]/{\rm
  H}\alpha)=-0.79\pm0.04$, and $\log([{\rm O I}] \lambda6300/{\rm
  H}\alpha)<-1.44$ ($3\sigma$ upper limit), which are consistent with
a star-forming galaxy based on the BPT diagrams
(Supplementary Figure~\ref{fig:bpt})\cite{baphte1981,veos1987}, as indicated by the
SDSS pipeline. The extinction-corrected H$\alpha$ luminosity is
$2.13\pm0.03\times10^{41}$ erg s$^{-1}$, which implies\cite{ke1998} a
star formation rate of $\sim$1.7 \msun/yr.

However, there seem to be some small residuals around the
[\ion{O}{III}] $\lambda$5007 line, indicating systematic blue
excess. Fitting the star light subtracted spectrum around the
[\ion{O}{III}] $\lambda$5007 region with two Gaussian lines, one for
the strong narrow component and the other for the broad wing, we infer
the wing to have $\sigma=555_{-161}^{+283}$ km s$^{-1}$ and to be
blue-shifted relative to the narrow line center by $312_{-233}^{+271}$
km s$^{-1}$. The flux of the blue wing is $1.3\pm0.4\times10^{-17}$
erg s$^{-1}$ cm$^{-2}$, corresponding to a 5.0$\sigma$
detection. Compared with the fitting residuals in other places, the
residuals here are about the strongest, and thus the blue wing could
be real, but we cannot completely rule out that it is due to enhanced
sky noise caused by a bright sky line near [\ion{O}{III}]
$\lambda$4959. A blue wing around [\ion{O}{III}] $\lambda$5007 is
commonly seen in AGNs and starburst galaxies and has been interpreted
as due to outflowing ionized gas\cite{koxuzh2008,
  mualfi2013,haalmu2014,ruvesa2005}. For XJ1500+0154, the blue wing
around [\ion{O}{III}] $\lambda$5007, if real, is more likely due to
star-forming activity, instead of due to the X-ray outburst, as the
nucleus is significantly obscured.

The mass and light (source rest-frame 3200--6800 \AA) distributions of
the stellar populations with respect to the age from the pPXF fit are
shown in Supplementary Figure~\ref{fig:s132740_agefraction}, indicating the presence
of some very young ($<5$ Myr) populations, although the mass is
dominated by old populations. The mean age weighted by light is 1.6
Gyr, and the mean age weighted by mass is 5.0 Gyr. The total stellar
mass is $6.4\times10^9$ \msun, and the total stellar luminosity is
$1.9\times10^9$ \lsun, suggesting a dwarf galaxy (comparable to the
Large Magellanic Cloud). Using the relation between the BH mass and
the total galaxy stellar mass\cite{revo2015}, we estimated the BH mass
to be $1.5\times10^6$ \msun\ (the $1\sigma$ uncertainty is 0.55 dex).

We do not see a clear broad component of the H$\alpha$ line. Some
TDEs, which would be the best interpretation for XJ1500+0154, did show
broad emission lines. The well covered TDE ASASSN-14li had peak
H$\alpha$ luminosity of $1.1\times10^{41}$ erg s$^{-1}$ with the
corresponding FWHM of $\sim$3000 km s$^{-1}$, detected in early
epochs\cite{hokopr2015}. Such an emission line, however, would be
difficult to detect for XJ1500+0154 if we assume the Galactic
reddening of $\mathrm{E(B-V)}_\mathrm{G}=0.05$ mag and intrinsic
reddening of $\mathrm{E(B-V)}_\mathrm{i}=0.71$ mag, which is based on
$\mathrm{E(B-V)}=1.7\times 10^{-22} N_\mathrm{H}$ and intrinsic column
density $N_{\rm H,i}=0.42\times10^{22}$ cm$^{-2}$ (inferred from our
X-ray spectral fits; see below). Therefore, we cannot rule out that a
broad H$\alpha$ line might be present in the source but is hard to
detect due to strong extinction. However, it is also possible that
XJ1500+0154 has intrinsically very weak optical emission (either broad
emission lines or the continuum) associated with the X-ray outburst,
as observed in some other TDEs during the X-ray bright
phase\cite{sarees2012} \\

\noindent \textbf{The source evolution with normalized count rates}

In order to study the source behavior in a relatively
model-independent way, we calculated the count rates normalized to
those expected based on the CompTT fit to the C3--C9
spectrum. Although the CompTT fit was used, other models that fit the
C3--C9 spectra reasonably well (e.g., an absorbed PL) gave very
similar results.

Supplementary Figure~\ref{fig:ltcr} shows the long-term evolution of such normalized
count rates in $b=$ 0.4--3 keV, $b1=$ 0.4--1 keV and $b2=$ 1--3
keV. The source appears to show a prolonged outburst. It was not
detected in C1 on April 2nd 2005, with the 0.4--3 keV normalized count
rate $R_\mathrm{N,b}<2.3$\% ($3\sigma$ upper limit, upper panel in
Supplementary Figure~\ref{fig:ltcr}). The source was detected less than 4 months
later in X1 on July 23rd 2005, with $R_\mathrm{N,b}=7.0\pm2.6$\% (90\%
error). It was detected much brighter three years later, with
$R_\mathrm{N,b}=1.9\pm0.1$ for C2 on June 5th 2008 and
$R_\mathrm{N,b}=2.3\pm0.1$ and $2.1\pm0.1$ for X2 and X3,
respectively, in February 2009. The source flux decayed very slowly
later, with $R_\mathrm{N,b}=1.01\pm0.02$ in C3--C9 in March--April
2011 and $0.35\pm0.04$ in C10 on February 23rd 2015. Similar
normalized count rates were also observed in the \emph{Swift}
monitoring observations in 2014--2016, though they have poor
statistics. Overall, the source count rate seemed to show a fast rise,
by a factor of $>$97 and probably in just months, and experienced a
very slow decay, by a factor of a few over a decade in time.

Comparing the 0.4--1 keV and 1--3 keV count rates (middle and lower
panels in Supplementary Figure~\ref{fig:ltcr}), we detected significant softening of
the source in C10, due to a significant drop in the 1--3 keV count
rate. The normalized 0.4--1 keV and 1--3 keV count rates of C10 are
$R_\mathrm{N,b1}=0.68\pm0.10$ and $R_\mathrm{N,b2}=0.15\pm0.04$, while
C3--C9 has $R_\mathrm{N,b1}=1.01\pm0.03$ and $1.02\pm0.03$ (close to
one by definition). The $R_\mathrm{N,b2}$ to $R_\mathrm{N,b1}$ ratios
are $0.21\pm0.06$ and $1.01\pm0.04$ (90\% error) for C10 and C3--C9,
respectively, indicating the softening of C10 with respect to C3--C9
at the 17.8$\sigma$ confidence level. In comparison, C2, X2, and X3
have the $R_\mathrm{N,b2}$ to $R_\mathrm{N,b1}$ ratios of
$0.90\pm0.10$, $0.90\pm0.07$, and $0.79\pm0.06$, respectively, close
to that of C3--C9. Other observations have poor statistics. S1 and
S6--S8 seems to be as bright and as hard as C3--C9. S2--S5 is fainter
and could be consistent with C10. X1, in the rise phase of the
outburst, could have a relatively hard spectrum too.  \\

\noindent \textbf{Spectral state identification for X2, X3, C2, and C3--C9}
  
Supplementary Table~\ref{tbl:spfitmcdpl} gives the fit results using simple models:
a PL, a MCD ({\it diskbb} in XSPEC), and their combination MCD+PL. The
spectra of X2, X3, C2, and C3--C9 have similar best-fitting spectral
parameters in these models, implying the same emission mechanism in
these observations. The absorbed PL fits gave the photon index
$\Gamma_\mathrm{PL}\sim4$--5 for X2, X3, and C2, and C3--C9,
indicating very soft spectra and ruling out that the source was in the
hard spectral state of an accreting BH in these observations. X2, X3,
and C2 are relatively close in time (within $\sim$8 months), and their
spectra are a little softer ($\Gamma_\mathrm{PL}\sim5$) than that of
C3--C9 ($\Gamma_\mathrm{PL}\sim4$), taken two years later after X2 and
X3. The X2, X3, and C2 spectra can be generally described with an
absorbed MCD, but the fit with an absorbed MCD to C3--C9 shows
systematic positive residuals at high energies $>$2 keV
(Supplementary Figure~\ref{fig:delchi}). Adding a PL (i.e., total model MCD+PL) with
$\Gamma_\mathrm{PL}$ fixed at a value of 2.5, typically seen in the
Galactic BHBs in the thermal state, improved the fit to C3--C9 most
significantly, but it still shows some systematic positive residuals
in 2--3 keV and negative ones above 3 keV
(Supplementary Figure~\ref{fig:delchi}). In any case, the MCD+PL fits inferred a
very hot disk of $kT_\mathrm{MCD}\sim0.3$ keV in all spectra (the
MCD+PL fits were not sensitive to the $\Gamma_\mathrm{PL}$ value
assumed, as long as $\Gamma_\mathrm{PL}$ was constrained to be
$\lesssim$3.5; for higher values of $\Gamma_\mathrm{PL}$, the fits
obtained would be dominated by a PL, instead of by a disk). We note
that all fits to X2 tested above in fact showed some systematic
residuals, whose origin will be discussed more later.

The MCD+PL fits are roughly statistically acceptable, but there are
problems to associate them with the thermal state, which is
characterized by a dominant thermal disk at the sub-Eddington
luminosity\cite{remc2006}. The inferred disk temperatures of
$kT_\mathrm{MCD}\sim 0.3$ keV are too high for such a state of an
accreting SMBH. For a standard thermal disk, the maximum temperature
$kT_\mathrm{MCD}\propto M^{-1/4}(L_\mathrm{MCD}/L_\mathrm{Edd})$. The
Galactic BH X-ray binaries have maximum disk temperatures
normally\cite{remc2006,dogiku2007} $\lesssim$1 keV. The MCD+PL fits
inferred $L_\mathrm{X}\sim1$--$2\times10^{43}$ erg
s$^{-1}$. Therefore, for XJ1500+0154 to be at the sub-Eddington limit
in the observations considered, the BH mass should be at least $10^5$
\msun. Then the maximum temperature $kT_\mathrm{MCD}$ should be
$\lesssim$0.1 keV, much lower than the values inferred.  The high disk
temperature might be possible if the SMBH considered is maximally
spinning. To test this further, we tried to fit X3, C2, C3--C9
simultaneously with the more physical AGN model
optxagnf\cite{dodaji2012} (in XSPEC). Because we are considering the
thermal state, in the optxagnf model, we assumed that the
gravitational energy released in the disk is emitted as a
color-corrected blackbody down to a (coronal) radius $r_\mathrm{cor}$,
while within this radius the available energy is released in a hard PL
form Comptonization component in an optically thin hot corona of
temperature 100 keV. The PL index was fixed at a value of 2.5. The BH
mass, the BH spin and the intrinsic column density was tied to be the
same but allowed to vary. The Eddington ratios were forced to be below
1.0 (thus sub-Eddington). With all these settings, the best fit
required a maximally spinning BH with mass $2.8\times10^5$ \msun, but
still a relatively high reduced $\chi^2$ value was obtained (1.35 for
280 degrees of freedom), and systematic fit residuals were seen
clearly.

Given the above problem of fitting the spectra with the standard
thermal state model MCD+PL, we tried to test the Comptonization model
CompTT\cite{ti1994} (in XSPEC). This model is commonly used to fit the
X-ray spectra of ULXs\cite{glrodo2009}, most of which are believed to
be the super-Eddington accreting stellar-mass BHs. We found that we
cannot constrain all parameters well, and it seems that the fits can
either assume cold seed photon solutions of $kT_0\lesssim0.1$ keV or
hot seed photon solutions of $kT_0\sim0.1$--0.2 keV, corresponding to
two local/global minima in $\chi^2$, which differed by $<$6. Cold seed
photon solutions were associated with higher column densities
($N_\mathrm{H}\sim0.4$--$0.5\times10^{22}$ cm$^{-2}$) than hot seed
photon solutions ($N_\mathrm{H}\sim0.1$--$0.2\times10^{22}$
cm$^{-2}$). In the case of ULXs, the seed photon temperatures were
often inferred\cite{glrodo2009} to be $\lesssim$0.3 keV. If our
spectra have the same emission mechanism as most ULXs and the seed
photons are from a thermal disk, we would expect the seed photons in
our spectra to be very cool, $<0.1$ keV. Therefore we will focus on
the cold seed photon solutions.  Still, we found that in order to
obtain well constrained corona temperatures and optical depths for
better comparison between observations, it is desirable to fix the
intrinsic column density $N_\mathrm{H,i}$ and the seed photon
temperature $kT_0$. From the simultaneous fits to X3, C2, and C3--C9
spectra with these parameters tied to be the same (X2 was not
included, again due to presence of fit residuals), we inferred
$N_\mathrm{H,i}=0.42\pm0.06\times10^{22}$ cm$^{-2}$ and
$kT_0=0.04\pm0.04$ keV. Therefore in the final fits with CompTT, we
fixed $N_\mathrm{H}$ at $0.42\times10^{22}$ cm$^{-2}$ and $kT_0$ at
$0.04$ keV. The fit results are given in Table~1. The
disk geometry was assumed (the fits of similar quality can also be
obtained assuming a spherical geometry, which would infer similar
corona temperatures but higher (by a factor of $\sim$2) values of the
optical depth).

From Table~1, we see that the CompTT fits suggest a cool
($kT_\mathrm{e}\sim0.35$--1.3 keV) optically thick ($\tau\sim4$--11)
corona for X2, X3, C2 and C3--C9. C3--C9 might have a little hotter
corona and a lower optical depth than earlier observations X2, X3, and
C2. To test the significance of the presence of a cool optically thick
corona in these spectra, we still tried to fit X3, C2, and C3--C9
simultaneously, with the seed photon temperatures all fixed at 0.04
keV, the corona temperatures all fixed at 20.0 keV, and the column
densities allowed to vary freely but all tied to be the same. Then we
obtained a total $\chi^2$ value of 257.6 for 282 degrees of
freedom. In comparison, the total $\chi^2$ value is 242.3 for 3 fewer
degrees of freedom if the corona temperature parameters are all
allowed to vary freely. Based on the $F$-test, this implies a
$3.4\sigma$ improvement of the fits with a cool optically thick corona
over those with a hot optically thin one, under the assumption of cool
seed photons. Therefore there is evidence that XJ1500+0154 was in a
super-Eddington accretion state in X2, X3, C2, and C3--C9.

We note that all the spectral models tested above on X2 all showed
systematic fit residuals, as shown in Supplementary Figure~\ref{fig:delchi}. X2 was
taken only six days before X3, and they have very similar instrument
configurations and source off-axis angles. The source spectral shape
and count rates are also very similar in these two observations, which
seems to suggest that the residuals in X2 could be due to some
calibration uncertainty, or due to presence of a warm absorber in
X2. Starting with the continuum model CompTT with $N_\mathrm{H,i}$
fixed at $0.42\times10^{22}$ cm$^{-2}$ and $kT_0$ fixed at $0.04$ keV,
we found that the residuals can be significantly reduced by adding an
edge of threshold energy $E_\mathrm{edge}=0.63\pm0.03$ keV and optical
depth $\tau=0.88\pm0.30$ (corresponding to a $5\sigma$ improvement)
and another edge of $E_\mathrm{edge}=1.33\pm0.07$ keV and
$\tau=0.37\pm0.21$ (corresponding to a $3\sigma$
improvement). Replacing these edges with the more physical model
zxipcf, we obtained two possible fits. One inferred an absorber of
column density $N_\mathrm{H}=0.4\times10^{22}$ cm$^{-2}$, ionizing
parameter $\log(\xi)=-0.55$ (i.e., nearly neutral), and zero speed
(i.e., the absorber is static relative to the source), and required
the unabsorbed luminosity to be a factor of 6.7 higher than that of
X3. The other fit inferred an absorber of
$N_\mathrm{H}=3.5\times10^{22}$ cm$^{-2}$, $\log(\xi)=2.7$ and
redshifted speed of 29\% the speed of the light (thus an inflow), with
the unabsorbed source luminosity similar to that of X3. Both fits
accounted for most of the residuals around 0.6 keV and are clearly
challenging to understand, with the former requiring a large change in
luminosity in a short time and the latter requiring a fast warm
inflow. Future observations would be helpful to determine whether the
residuals are real. \\

\noindent \textbf{Spectral state identification for C10}

We also have a relatively good spectrum from C10, which appears to be
much softer than earlier observations, as shown above based on the
normalized count rates in different energy bands. The absorbed PL fit
inferred a very high photon index of $\Gamma_\mathrm{PL}=8.0\pm1.2$
for C10, compared with $\Gamma_\mathrm{PL}\sim4$--5 for X2, X3, C2 and
C3--C9. The MCD+PL fit with $\Gamma_\mathrm{PL}$ fixed at a value 2.5
also indicates that the spectrum in C10 is dominated by a disk with a
much lower temperature in C10 than in X2, X3, C2 and C3--C9 (0.13 keV
versus $\sim$0.3 keV).

One explanation for the spectral softening in C10 is that the source
showed a transition from the super-Eddington accretion state in
earlier observations to the thermal state in C10. Then the MCD+PL fit
would be a reasonable model. Table~1 lists such a fit
with $N_\mathrm{H,i}$ fixed at $4.2\times10^{21}$ cm$^{-2}$, as
obtained from the simultaneous CompTT fits to X3, C2, and C3--C9. This
model inferred $L_\mathrm{X}$ in C10 only lower than that in C3--C9 by
15\%. We note that in the second ULX in M31, similar state transition
also occurred with luminosity changing\cite{mimima2013} by $<$20\%.

Alternatively, the spectral softening in C10 could be due to a
transient warm absorber in C10 that obscured the high-energy flux. To
test this scenario, we added a warm absorber to the CompTT fit to
C3--C9. The warm absorber was modelled with the \texttt{zxipcf} (in
XSPEC) model. The CompTT parameters were fixed at the values obtained
from the fit to C3--C9.  The covering fraction was found to be
consistent with 1.0 (the 90\% lower confidence bound is 0.91), and we
thus fixed it at the value of 1.0. Then we inferred the warm absorber
to have $N_\mathrm{H}=64\times10^{22}$ cm$^{-2}$, $\log(\xi)=2.8$ and
blueshifted velocity of 0.36c (Table~1). This fit has
almost the same $C$ statistic as the MCD+PL fit. If the normalization
of the CompTT model is allowed to vary, considering that the source
might become fainter, we found that the unabsorbed luminosity in C10
should be $>$88\% (90\% lower bound, the upper bound cannot be
constrained, as there is degeneracy between the source flux and the
column density of the absorber) of that of C3--C9. A highly ionized
sub-relativistic outflowing absorber is expected in the
super-Eddington accretion phase of TDEs\cite{re1988} and had been
inferred in a few cases\cite{albegu2016,limair2015}. \\

\noindent \textbf{Spectral state identification for Swift observations}

The \emph{Swift} observations are all too faint to carry out detailed
spectral fits. In S1, S2--S5, and S6--S8, we collected 18.3, 6.0 and
18.8 net counts within 0.3--10 keV in exposures of 8.7 ks, 9.0 ks, and
6.7 ks, respectively. Based on the normalized count rates shown in
Supplementary Figure~\ref{fig:ltcr}, the source spectra in S1 and
S6--S8 are more consistent with that in C3--C9 than with that in
C10. The normalized 1--3 keV count rates in S1 and S6--S8 are both
consistent with that in C3--C9 to within 1$\sigma$ error but both
different from that in C10 at the $2.5\sigma$ confidence
level. Therefore the source might be still in the super-Eddington
accretion state in S1 and S6--S8. In contrast, the low 0.4--3 keV
normalized count rate in S2--S5 makes it more consistent with C10
(within $1\sigma$) than C3--C9, and the source could be in the thermal
state or subject to transient absorption again in this
observation. Given the low statistics of the \emph{Swift}
observations, however, we cannot completely rule out that the source
had similar X-ray spectra in S2--S8 as in C10, which could suggest
that the source had been completely settled to the thermal state since
C10. Deeper observations of the source in the future are needed to
firmly determine the duration of the super-Eddington accretion
phase.\\

\noindent \textbf{Spectral state identification for X1}

The spectrum of X1 has low quality too. Fitting it with a PL gave
$\Gamma_\mathrm{PL}=2.7_{-0.7}^{+2.3}$. Thus it could be a little harder than
other observations.  We tried to fit it with the MCD+PL model with
$N_\mathrm{H,i}$ fixed at $4.2\times10^{21}$ cm$^{-2}$ and the PL
photon index fixed at the value of 2.5. We obtained a fit with a cool
disk of $kT_\mathrm{MCD}=0.10_{-0.04}^{+0.06}$ keV and a PL
contributing about 25\% of the rest-frame 0.34--11.5 keV luminosity.

The identification of the spectral state for X1 is clearly non-trivial
due to its low statistics, but it is consistent with a transitional
state, as expected considering that it was at the beginning of the
outburst.\\

\noindent \textbf{Fitting the long-term X-ray luminosity curve}

The evolution of the X-ray luminosity in TDEs depends on many factors
and can involve physics that is still poorly understood, especially in
the super-Eddington accretion regime. We explored the models that
could describe the data assuming full disruption of a star by a $10^6$
\msun\ black hole. The accretion rate of the mass was assumed to be
slowed relative to the fallback rate by the viscous timescale
$\tau_{\rm visc}$, according to the equation
\begin{equation}
\dot{M}_{\rm d}(t) = \frac{1}{\tau_{\rm visc}} \left(e^{-t/\tau_{\rm visc}} \int_{0}^{t} e^{t^{\prime}/\tau_{\rm visc}} \dot{M}_{\rm fb}(t^{\prime}) dt^{\prime} \right),
\end{equation}
where $\dot{M}_{\rm d}$ is the mass accretion rate and $\dot{M}_{\rm
  fb}$ is the mass fallback rate\cite{gura2013}.  The radiative
  efficiency was assumed to be 0.1 when the accretion rate $\dot{M}_{\rm d}$
  is below the 0.5 isotropic Eddington limit
  $\dot{M}_\mathrm{Edd}$. At higher accretion rates, the inner disk
  could begin to reach the local Eddington limit\cite{lireho2009}, and
  we assumed the luminosity to scale logarithmically, in the
  form\cite{kimu2016} of
  $1.0+\log(\dot{M_\mathrm{d}}/0.5\dot{M}_\mathrm{Edd})$. Based on
  the X-ray spectral fits, we assumed that 25\% of the light was
  emitted in the X-ray band (the source rest-frame 0.34--11.5 keV).

Supplementary Figure~\ref{fig:lumlcmodel} shows the expected light curves for
several models of different masses of the disrupted star and different
viscous timescales. For comparison, models incorporating the
super-Eddington accretion effects are shown as thick black lines,
while those not are shown as thin gray lines (representing the
evolution of the fallback rate in case of prompt circularization).

We find that there is degeneracy between the stellar mass and
$\tau_{\rm visc}$. Our data can be described with the full disruption
of a 2 \msun\ star with $\tau_{\rm visc} = 3$~yr (Figure 2) or a 10
\msun\ star with $\tau_{\rm visc} = 0$
(Supplementary Figure~\ref{fig:lumlcmodel}). The degeneracy can be broken with more
future monitorings of the source, as the luminosity is expected to
decay faster for disruption of a lower mass star. Assuming a higher
efficiency in either the Eddington excess or in the mass-to-light
conversion would adjust the favored stellar masses downward, with a
maximally-spinning black hole driving the favored masses to values of
0.5 and 2.5 $M_{\odot}$, respectively.

While the case with a larger star describes our data just as well as
the case with a star of nearly a solar mass, the steep stellar mass
function strongly favors the disruptions of lower-mass stars, even in
cases where a recent starburst has occurred\cite{ko2016}, with $< 1$\%
of all disruptions coming from stars with $M \geq 10$
\msun. Additionally, we expect that disruptions of stars around
$10^{6} M_{\odot}$ black holes will be slowed by viscous effects in
the majority of disruptions, with two-thirds of disruptions having
viscous timescales longer\cite{gura2015} than a year for $M_{\rm h} <
10^{6} M_{\odot}$. Because of these expectations, we argue that
XJ1500+0154 most likely originated from the disruption of a nearly
solar-mass star.

A single strong disruption and prompt accretion of an evolved star,
either ascending the red giant branch or on the horizontal branch, can
produce a prolonged TDE with a fast rise and a prolonged
super-Eddington accretion phase\cite{magura2012}, as seen in
XJ1500+0154. However, TDEs of evolved stars are expected to be very
rare for a black hole of mass $10^6$ \msun, accounting for $\sim$1\%
of the total TDEs\cite{ko2016}, and they should be dominated by
repeated, partial disruption\cite{maragr2013}. Single strong
disruptions of evolved stars should be extremely rare, and we do not
expect to detect any in the limited volumn of space searched (see
below).\\

\noindent \textbf{ROSAT Observations}

XJ1500+0154 was not detected in the {\it ROSAT} All-Sky Survey in
1990. The survey had a detection limit\cite{voasbo1999} of the
0.1--2.4 keV flux of 5$\times$10$^{-13}$ erg s$^{-1}$ cm$^{-2}$, which
is a factor of 3 higher than the peak absorbed 0.1--2.4 keV
(observer-frame) flux of XJ1500+0154 ($\sim$$1.7\times10^{-13}$ erg
s$^{-1}$ cm$^{-2}$ in X2 and X3). The source was not detected either
in the {\it ROSAT}/HRI pointed observation in 1998. The $3\sigma$
upper limit of the source net count rate is 0.0038 counts s$^{-1}$,
which is about 50\% lower than that expected from X2 and
X3. Therefore, although the source was not detected in \emph{ROSAT}
observations, they are not deep enough for us to rule out the emission
level seen in the current outburst.\\

\noindent \textbf{Short-term X-ray variability}

Supplementary Figure~\ref{fig:srclc} shows the light curves for all bright
observations (X2--X3 and C2--C10). We obtained $\chi^2$ probability of
constancy of 0.13 and $2.9\times10^{-4}$ from the 0.3--3 keV pn light
curves binned at 500 s in the X2 and X3 observations,
respectively. Therefore, no short-term variability was detected from
X2. X3 might show some variability, with a fast (within 2 ks) drop to
almost zero count rate at around 20 ks into the observation. The
average count rates of X2 and X3, which are six days apart, are,
however, consistent with each other within $2\sigma$
(Supplementary Table~\ref{tbl:obslog}). We detected no short-term variability in the
\emph{Chandra} observations C2--C10, all with a Gregory-Loredo
variability index of 0. We note that the lower count rates in the {\it
  Chandra} observations (a factor of $\gtrsim$2 lower than the
\emph{XMM-Newton}/pn in X2 and X3) could make it hard to detect the
fast drop as is possibly present in X3. The observations C3--C9, taken
within a 13 day period, have count rates consistent with each other
within $2\sigma$ (Supplementary Table~\ref{tbl:obslog}).\\

\noindent \textbf{UV and optical long-term variability}

The SDSS photometry of the host galaxy on May 23rd 2001 is
$u\arcmin=21.71\pm0.35$ AB mag, $g\arcmin=20.96\pm0.07$ AB mag,
$r\arcmin=20.34\pm0.06$ AB mag, $i\arcmin=19.97\pm0.07$ AB mag, and
$z\arcmin=19.81\pm0.27$ AB mag (Petrosian magnitudes). The
corresponding CFHT/MegaPrime $r\arcmin$ and $g\arcmin$ photometry can
be obtained through the relation $r\arcmin_\mathrm{Mega} =
r\arcmin_\mathrm{SDSS} - 0.024 (g\arcmin_\mathrm{SDSS} -
r\arcmin_\mathrm{SDSS})=20.32\pm0.06$ AB mag, and
$g\arcmin_\mathrm{Mega} = g\arcmin_\mathrm{SDSS} - 0.153
(g\arcmin_\mathrm{SDSS} - r\arcmin_\mathrm{SDSS})=20.87\pm0.07$ AB
mag. In comparison, we measured $r\arcmin_\mathrm{Mega}=20.25\pm0.01$
AB mag and $g\arcmin_\mathrm{Mega}=20.92\pm0.01$ AB mag from the
CFHT/MegaPrime stacked images in 2014. Therefore the photometry in
each filter was consistent with each other in these two epochs to
within the $1\sigma$ uncertainty. The $3\sigma$ upper limit in the
$r\arcmin$ band in 2014 due to the flare is about $\lambda
L_\lambda<1.3\times10^{42}$ erg s$^{-1}$ or $<3.5\times10^8$ \lsun. If
we adopt the reddening relation $\mathrm{E(B-V)}=1.7\times 10^{-22}
N_\mathrm{H}$ and use the column density of
$N_\mathrm{H,i}=4.2\times10^{21}$ cm$^{-2}$ inferred from the X-ray
spectral fits, the above limit after extinction correction would be an
order of magnitude larger (i.e., $<3.5\times10^9$ \lsun). The three
ASASSN TDEs have $\lambda L_\lambda$ in the range between $10^9$ and
$10^{10}$ \lsun\ at $\lambda=5400\AA$ (the source rest-frame central
wavelength of the $r\arcmin$ band for XJ1500+0154) at the very early
epoch of the events\cite{hokopr2016}. Therefore we cannot rule out a
similar level of optical emission as seen in the ASASSN TDEs in our
event.

There is a faint UV source from the \textit{Swift} observation at the
position of the SDSS galaxy. We obtained the W1, M2, and W2 magnitudes
of $23.1\pm0.5$ AB mag, $23.3\pm0.5$ AB mag, and $23.8\pm0.7$ AB mag,
respectively, from S1, and $23.5\pm1.4$ AB mag, $22.8\pm0.3$ AB mag,
and $22.9\pm0.3$ AB mag, respectively, from S2--S8. The relatively
large errors are due to the scattered light background from a nearby
bright star. The UV source was also detected by GALEX on June 3rd, 2007,
with the NUV and FUV magnitudes of $22.9\pm0.3$ AB mag and
$23.8\pm0.3$ AB mag (the NUV filter has an effective wavelength
similar to that of the \textit{Swift} M2 filter, i.e., $\sim$2230
\AA), The blue UV emission is consistent with each other in these
epochs and is consistent with that expected from emission from the
young stellar populations based on the pPXF fit to the SDSS spectrum
of the galaxy.  The UV emission from the three ASASSN TDEs in the
early epoch\cite{hokopr2016} would correspond to a W2 magnitude of
$>$27 AB mag for XJ1500+0154 if we also applied the extinction based on
$\mathrm{E(B-V)}=1.7\times 10^{-22} N_\mathrm{H}$ and used the column
density of $N_\mathrm{H,i}=4.2\times10^{21}$ cm$^{-2}$. The U band
magnitude in S2--S8 ($22.1\pm0.3$ AB mag) is also consistent with the
stellar emission.\\

\noindent \textbf{Arguments against the AGN explanation}

The coincidence with an optical galactic center, with chance
coincidence probability of only 0.000017, led us to conclude that
XJ1500+0154 is due to nuclear activity in SDSS
J150052.07+015453.8. The significant absorption column density other
than the Galactic value required in the X-ray spectral fits reassures
this association. Then we are left with two possible interpretations
of the source: a persistent AGN or a TDE.

As we have shown, we see no sign of persistent nuclear activity in the
optical spectrum, which shows no clear broad emission lines but narrow
ones, with the ratios fully consistent with those expected for
starburst galaxies. For AGNs, there is a strong correlation between the
persistent hard X-ray luminosity and the extinction-corrected [\ion{O}{III}]
$\lambda$5007 luminosity $L^{\rm c}_{\rm OIII}$. XJ1500+0154 has
$L^{\rm c}_{\rm OIII}=2.9\pm0.1\times10^{41}$ erg s$^{-1}$ if we
assume $\mathrm{E(B-V)}_\mathrm{G}=0.05$ mag and
$\mathrm{E(B-V)}_\mathrm{i}=0.71$ mag, which is again based on
$\mathrm{E(B-V)}=1.7\times 10^{-22} N_\mathrm{H}$ and intrinsic column
density $N_{\rm H,i}=0.42\times10^{22}$ cm$^{-2}$ inferred from our
X-ray spectral fits. The 2--10 keV luminosity corresponding to this
$L^{\rm c}_{\rm OIII}$ is\cite{labima2009} $3.6\times10^{42}$ erg
s$^{-1}$ (the dispersion is 0.63 dex). The corresponding 0.34--2.0 keV
unabsorbed luminosity assuming a PL of photon index 2.0 is
$3.9\times10^{42}$ erg s$^{-1}$, which is about an order of magnitude
lower than the flux seen in the outburst. Given that the line ratios
are fully consistent with the star-forming activity, the
[\ion{O}{III}] $\lambda$5007 flux should be dominated by the
star-forming activity, instead of that of the nucleus. Then the
persistent unabsorbed 0.34--2 keV luminosity implied by the optical
spectrum would be much lower than observed in the ourburst, arguing
against a persistent AGN explanation for XJ1500+0154.

We can also compare XJ1500+0154 with 753 spectroscopically identified
AGNs in Lin et al.\cite{liweba2012}. These AGNs were included in that
study because they had multiple observations, with at least one
detection with S/N $>$ 20, in the 2XMM-DR3
catalog. Supplementary Figure~\ref{fig:agnxraycolor} shows an X-ray color-color
diagram, adapted from Figure~2 in Lin et al., by using the 3XMM-DR5
catalog. The X-ray colors HR2 and HR3 are defined as $(H-S)/(H+S)$,
with $S$ and $H$ being the MOS1-medium-filter equivalent 0.5--1 keV
and 1--2 keV counts rates for HR2 and 1--2 keV and 2--4.5 keV count
rates for HR3, respectively. The MOS1-medium-filter equivalent count
rates are those expected for an on-axis MOS1 observation using a
``Medium'' optical filter\cite{liweba2012}. In terms of HR3, which
characterizes the spectral shape within 1--4.5 keV, XJ1500+0154 is
significantly softer (HR3 in the range between $-0.78$ and $-0.86$ for
all bright observations C2, C3--C9, C10, X2 and X3) than AGNs (HR3
$\gtrsim$ $-0.70$).

The long-term variability of XJ1500+0154, a factor of $>$97 ($3\sigma$
lower limit), is also extreme, compared with AGNs. Only one out of 753
AGNs in Lin et al. varied by a factor of $>$97 based on the 3XMM-DR5
catalog. The large variability in AGNs has been normally ascribed to
all kinds of absorbers\cite{grkosc2013}. In order to explain the
non-detection of XJ1500+0154 in C1 and the low count rate in X1, it
would require a neutral absorber of $N_\mathrm{H}>6.8\times10^{22}$
cm$^{-2}$ and $N_\mathrm{H}=2.5\times10^{22}$ cm$^{-2}$ fully covering
a X2 spectrum in C1 and X1, respectively. The problem with this
explanation is that it would imply no detection of the source below 1
keV in X1, while we had detected the source in 0.4--1 keV at the
$3.2\sigma$ confidence level in this observation, unless the absorber
has a complex structure. Besides, this explanation has to require
highly variable absorption of an AGN with unusually soft X-ray
spectra.

A very small number of high-amplitude very soft X-ray flares were
detected from galactic nuclei that might have low persistent luminosities as
inferred from the optical emission
lines\cite{tekaaw2012,hokite2012,liirgo2013,liweba2014,misaro2013,samoko2015,grkosa2015,camaco2015}. It
is under debate on whether these flares are due to, e.g., disk
instability in AGNs, or due to TDEs. They might even represent a mixed
class, as they showed a variety of spectral shape and temporary
evolution. XJ1500+0154 is different from them in several aspects
(e.g., generally higher characteristic temperatures and dramatic
spectral softening in XJ1500+0154). Therefore it is not clear whether
XJ1500+0154 has the same origin as the other flares.

There are several AGNs known to show significant spectral changes,
between type 1 and type 2 in
optical\cite{shprgr2014,lacamo2015,pakoko2016} and/or between
Compton-thin and reflection-dominated states in
X-rays\cite{risael2009,rimiel2009}. There are various explanations for
them, including variable absorption and changes in the accretion rate
(due to, e.g., a TDE). Supplementary Figure~\ref{fig:agnxraycolor} in fact includes
some such changing-look AGNs, e.g., NGC 1365 and 1H
0707-495. These
AGNs change between two AGN standard spectral states with hard X-ray
spectra. XJ1500+0154 is different from them in that it changed between
two states unseen in standard AGNs, with very soft X-ray spectra.

Dwarf starburst galaxies hosting luminous AGNs are also extremely rare
in the local Universe.  Only a small number of dwarf galaxies are
known to show active
nuclei\cite{fisa1989,bahoru2004,resijo2011,pagogr2016}. Among them,
only Henize 2--10 is in a starburst galaxy, with extremely low
luminosity in the nucleus\cite{resijo2011}.

The column density inferred from the X-ray spectral fits is optically
thin to Thomson electron scattering\cite{fa1999}. Then the wind from
the central source can sweep up the gas into a shell and push it
outwards at a velocity\cite{sire1998} of
$v_s=(f_wL8\pi^2G/f_\mathrm{gas}\sigma^2)^{1/3}=2700(f_\mathrm{w}/f_\mathrm{gas})^{1/3}$
km s$^{-1}$, where $f_\mathrm{gas}$ is the mass fraction of the gas,
$\sigma=52$ km s$^{-1}$ is the velocity dispersion, $G$ is
gravitational constant, $L$ is the source bolometric radiation
luminosity, $\sim$$10^{44}$ erg s$^{-1}$, and $f_\mathrm{w}L$ is the
wind energy. Given the super-Eddington accretion nature of the source,
we expect $f_\mathrm{w}/f_\mathrm{gas}$ to be $>1.0$. Therefore
$v_s>2700$ km s$^{-1}$, and gas within 1 kpc (the effective radius of
the host galaxy) can be swept out in $<$0.4 Myr, which is very
short. Therefore the persistent luminous AGN explanation is hard to be
reconciled with the star-forming activity of the host galaxy.

We note that we do not have deep radio observations of XJ1500+0154 yet
and that it was not detected in the radio surveys NVSS or FIRST. The low
sensitivity of these surveys did not allow us to rule out a persistent
AGN based on the radio upper limit.\\

\noindent \textbf{Comparison with other candidate TDEs}

There are about 30 TDEs discovered thus far that are bright in
X-rays\cite{ko2012,ko2015}. Most of them had peak X-ray luminosity
$<$$10^{44}$ ergs s$^{-1}$, fast decay by about one order of magnitude
in a year, and super-soft X-ray spectra ($kT \lesssim
0.1$ keV)\cite{grthle1999,kogr1999,koba1999,essako2008,licagr2011,mauler2010,mauler2013,maliir2014,doceco2014,limair2015,mikami2015}. Figure
2 plots a dashed line showing a representative light curve for such
super-soft TDEs. Our event, evolving on a timescale at least two orders
of magnitude longer, having much harder X-ray spectra, and exhibiting
dramatic spectral softening, is different from them.

Recently, three TDEs with hard X-ray spectra (photon index around 2.0)
and apparent peak luminosity highly super-Eddington were also
found\cite{blgime2011,bukegh2011,zabeso2011,cekrho2012,brlest2015}. They
are also short-lived, decreasing by more than two orders of magnitude
in the first year. Our event is distinguished from them, with much
softer X-ray spectra, long duration, and dramatic X-ray spectral
change. \\

\noindent \textbf{Event rate}

We roughly estimated the rate of TDEs with a super-Eddington accretion
phase, characterized by quasi-soft X-ray spectra. XJ1500+0154 was
discovered through the search over the detections in the 3XMM-DR5
catalog that have S/N $>$ 20. We selected out sources that showed
large variability and/or soft spectra (e.g., HR3 $<-0.7$). Based on
observations outside the Galactic plane (the galactic latitude
$|b|>20^\circ$) and assuming that events like XJ1500+0154 have a mean
luminosity three times lower than that in X2 but have similar
quasi-soft X-ray spectra and last for 10 years, we estimated that we
should be able to detect about $1.8\times10^8 rn$ events, where $r$ is
the event rate per galaxy per year and $n$ is the galaxy density per
Mpc$^3$. We assumed\cite{dobrer2002} $n=1.4\times10^{-2}$
Mpc$^{-3}$. We only discovered one with quasi-soft X-ray spectra,
indicating $r\sim4\times10^{-7}$ gal$^{-1}$ yr$^{-1}$. This is a
conservative estimate because our search was still preliminary and the
absorption effect was not taken into account. The real rate could be a
factor of a few larger.

The above event rate calculation depends on the assumption of the
duration of the super-Eddington accretion phase in the
events. Assuming shorter durations would proportionally infer higher
rates. Therefore we cannot rule out a much higher rate for all
super-Eddington accreting TDEs as their super-Eddington accretion
phase could be generally shorter than that of our event. Although
super-Eddington accretion in TDEs is probably rarer than previously
thought, due to slow circularization effects, a significant fraction
of TDEs are still expected to show a super-Eddington accretion
phase\cite{gura2015}.  Guided by the discovery of XJ1500+0154, we
tried to search for TDEs with quasi-soft X-ray spectra (thus
suggesting a super-Eddington accretion phase) from newly public
\textit{XMM-Newton} observations and have discovered two candidates
(Lin et al. 2017, in preparation). They have peak X-ray luminosity of
$\sim$$10^{44}$ erg s$^{-1}$. One of them is clearly in an outburst,
with a rising time less than 10 months, and the host galaxy showed no
optical emission lines, ruling out the presence of a persistently
bright nucleus. The other one is consistent to be in an outburst too,
with two X-ray observations showing the flux decay by a factor of
$\sim$6 in 1.5 years, and the host is a star-forming galaxy (or at the
end phase of a starburst phase, given the strong Balmer absorption).
The durations of these two events still need to be constrained with
future observations. Discovery of additional quasi-soft X-ray objects
in bright outbursts and in hosts showing no sign of persistent nuclear
activity in the optical spectra strongly support that they are
super-Eddington accreting TDEs and that such events might not be so
rare.

\clearpage

\begin{table*}
\centering
\caption{\textbf{The X-ray Observation Log}. Columns: (1) the observation ID with our designation given in parentheses, (2) the observation start date, (3) the detector, (4) the off-axis angle, (5) the exposures of data used in final analysis, (6) radius of the source extraction region, (7) the net count rate in the source extraction region (0.3--3 keV for \emph{XMM-Newton} and 0.4--3 keV for \emph{Chandra} and \emph{Swift} observations), with 1$\sigma$ error (but for observations rh601125n00, C1 and S2--S5, the $3\sigma$ confidence bounds are given).}
\label{tbl:obslog}
\bigskip
\scriptsize
\sffamily
\begin{tabular}{rcccccccc}
\hline
\hline
Obs. ID &Date & Detector &OAA &$T$ (ks) &$r_{\rm src}$  & Count rate (10$^{-3}$ counts~s$^{-1}$)\\
(1) & (2) &(3) & (4) & (5) & (6) & (7)\\
\hline
\multicolumn{4}{l}{\emph{ROSAT}:}\\
rh601125n00 & 1998-01-18 & HRI & 13\farcm8 & 1.9 & 10$\arcsec$ & $0^{+3.8}$\\
\hline
\multicolumn{4}{l}{\emph{XMM-Newton}:}\\
\hline
0302460101(X1) &2005-07-23 & pn/M1/M2 & 13\farcm5 & 22/33/33 & 20$\arcsec$/20$\arcsec$/20$\arcsec$ &$1.45\pm0.40$/$0.74\pm0.20$/$0.21\pm0.15$\\
0554680201(X2) &2009-02-12 & pn/M2 & 12\farcm5 & 43/64 & 20$\arcsec$/20$\arcsec$ &$37.4\pm1.0$/$11.1\pm0.4$\\
0554680301(X3) &2009-02-18 & pn/M2 & 12\farcm5 & 42/64 & 20$\arcsec$/20$\arcsec$ &$35.6\pm0.9$/$9.7\pm0.4$\\
\hline
\multicolumn{4}{l}{\emph{Chandra}:}\\
\hline
5907(C1) & 2005-04-02 & ACIS-S1 &  12\farcm4 & 48 & 12\farcs4 & $0^{+0.28}$\\
9517(C2) & 2008-06-05 & ACIS-I3 &  14\farcm6 & 99 & 16\farcs7 & $11.1\pm0.3$\\
12951(C3) & 2011-03-28 & ACIS-S1 &  10\farcm9 & 74 & 13\farcs4 & $13.8\pm0.4$\\
13246(C4) & 2011-03-30 & ACIS-S1 &  10\farcm9 & 45 & 13\farcs4 &$13.9\pm0.6$\\
13247(C5) & 2011-03-31 & ACIS-S1 &  10\farcm9 & 36 & 13\farcs4 &$14.2\pm0.6$\\
12952(C6) & 2011-04-05 & ACIS-S1 &  10\farcm9 & 143 & 13\farcs4 & $13.4\pm0.3$\\
12953(C7) & 2011-04-07 & ACIS-S1 &  10\farcm9 & 32 & 13\farcs4 &$15.3\pm0.7$\\
13253(C8) & 2011-04-08 & ACIS-S1 &  10\farcm9 & 118 & 13\farcs4 &$13.8\pm0.3$\\
13255(C9) & 2011-04-10 & ACIS-S1 &  10\farcm9 & 43 & 13\farcs4 &$13.3\pm0.6$\\
17019(C10) &2015-02-23 & ACIS-S3 & 0\farcm3 & 37 & 1\farcs6 & $4.9\pm0.4$\\
\hline
\multicolumn{4}{l}{\emph{Swift}:}\\
\hline
00033207001(S1) & 2014-03-28 & XRT & 1\farcm0 & 8.7 & 25$\arcsec$ & $1.9\pm0.5$\\
00033207002(S2) & 2016-02-03 & XRT & 3\farcm8 & 3.1 & 25$\arcsec$ & \multirow{4}{*}{$0.7^{+1.3}$} \\
00033207003(S3) & 2016-02-04 & XRT & 2\farcm0 & 0.3 & 25$\arcsec$ &\\
00033207004(S4) & 2016-02-05 & XRT & 1\farcm5 & 2.7 & 25$\arcsec$ &\\
00033207005(S5) & 2016-02-07 & XRT & 1\farcm3 & 2.9 & 25$\arcsec$ &\\
00033207007(S6) & 2016-02-10 & XRT & 1\farcm2 & 3.0 & 25$\arcsec$ &\multirow{3}{*}{$2.8\pm0.7$}\\
00033207008(S7) & 2016-02-12 & XRT & 1\farcm0 & 2.3 & 25$\arcsec$ &\\
00033207009(S8) & 2016-02-14 & XRT & 1\farcm4 & 1.4 & 25$\arcsec$ &\\
\hline
\hline
\end{tabular}
\end{table*}

\clearpage

\begin{table*}
\centering
\caption{\textbf{Spectral fit results.} Columns : (1) the observations; (2) the model; (3) intrinsic absorption column density; (4) the apparent inner disk temperature of the MCD; (5) the MCD normization $N_{\rm MCD}\equiv ((R_{\rm MCD}/{\rm km})/(D/{\rm 10 kpc}))^2\cos\theta$, where $R_{\rm MCD}$ is the apparent inner disk radius, $D$ is the source distance, $\theta$ is the disk inclination; (6) the PL photon index; (7) the PL normalization; (8) the reduced $\chi^2$ and degrees of freedom for fits using the $\chi^2$ statistic (we rebinned the spectra to have at least 20 counts in each bin in order to use the $\chi^2$ statistic; other fits without reduced $\chi^2$ used the C statistic); (9) the 0.34--11.5 keV luminosity, corrected for the Galactic absorption but not instrinsic absorption (10) the 0.34--11.5 keV luminosity (the rest frame), corrected for the Galactic absorption and intrinsic absorption. The fits used spectra within 0.3--10 keV for \textit{XMM-Newton} and \textit{Swift} and used spectra within 0.3--7 keV for \textit{Chandra}. Parameters without errors were fixed in the fits. All errors are at the 90\%-confidence level.}
\label{tbl:spfitmcdpl}
\bigskip
\scriptsize
\sffamily
\begin{tabular}{r|c|ccccc|c|cc}
\hline
\hline
Observation & model &$N_{\rm H}$ & $kT_{\rm MCD}$ &$N_\mathrm{MCD}$ &$\Gamma_{\rm PL}$& $N_{\rm PL}$  & $\chi^2_\nu(\nu)$ & $L_{\rm abs}$ & $L_{\rm unabs}$\\
             &  & (10$^{22}$ cm$^{-2}$) & (keV) &&&($10^{-6}$)     &    & \multicolumn{2}{c}{(10$^{43}$ erg s$^{-1}$)}\\
(1)&(2)&(3)&(4)&(5)&(6)& (7)&(8)& (9)&(10) \\
\hline
\multirow{3}{*}{X1}&MCD+PL& $ 0.42$& $ 0.10^{+ 0.06}_{-0.04}$ & $210^{+27841}_{-204}$ & $2.5$& $  4.6^{+  3.9}_{ -3.4}$ &... & $ 0.08^{+ 0.04}_{-0.03}$  & $ 0.55^{+ 0.95}_{-0.31}$ \\
&MCD& $ 0.00^{+ 0.17}_{ 0.00}$ & $ 0.32^{+ 0.15}_{-0.11}$ & $ 0.08^{+ 0.93}_{-0.07}$ &...&...&... & $ 0.06^{+ 0.02}_{-0.02}$  & $ 0.06^{+ 0.05}_{-0.02}$ \\
&PL& $ 0.01^{+ 0.40}_{-0.01}$ &...&...& $ 2.71^{+ 2.34}_{-0.68}$ & $  4.4^{+  7.2}_{ -1.8}$ &... & $ 0.08^{+ 0.04}_{-0.04}$  & $ 0.08^{+ 0.51}_{-0.03}$ \\
\cline{1-10}
\multirow{3}{*}{X2} &MCD+PL& $ 0.24^{+ 0.05}_{-0.05}$ & $ 0.29^{+ 0.03}_{-0.03}$ & $ 5.67^{+ 4.11}_{-2.25}$ & $2.5$& $ 14.8^{+ 11.4}_{-11.6}$ &$1.03(105)$ & $ 1.22^{+ 0.07}_{-0.07}$  & $ 2.66^{+ 0.51}_{-0.42}$ \\
&MCD& $ 0.21^{+ 0.05}_{-0.04}$ & $ 0.31^{+ 0.02}_{-0.02}$ & $ 3.94^{+ 1.95}_{-1.27}$ &...&...&$1.06(106)$ & $ 1.17^{+ 0.06}_{-0.06}$  & $ 2.31^{+ 0.34}_{-0.27}$ \\
&PL& $ 0.64^{+ 0.08}_{-0.07}$ &...&...& $ 4.93^{+ 0.30}_{-0.28}$ & $464^{+ 94}_{-71}$ &$1.17(106)$ & $ 1.19^{+ 0.06}_{-0.06}$  & $25^{+12}_{-7}$ \\
\hline
\multirow{3}{*}{X3} &MCD+PL& $ 0.20^{+ 0.05}_{-0.05}$ & $ 0.28^{+ 0.03}_{-0.03}$ & $ 4.93^{+ 4.19}_{-2.09}$ & $2.5$& $ 19.3^{+ 11.7}_{-11.9}$ &$0.88( 98)$ & $ 1.19^{+ 0.07}_{-0.07}$  & $ 2.35^{+ 0.47}_{-0.38}$ \\
&MCD& $ 0.16^{+ 0.04}_{-0.04}$ & $ 0.32^{+ 0.02}_{-0.02}$ & $ 2.91^{+ 1.49}_{-0.96}$ &...&...&$0.94( 99)$ & $ 1.12^{+ 0.05}_{-0.05}$  & $ 1.93^{+ 0.27}_{-0.22}$ \\
&PL& $ 0.55^{+ 0.07}_{-0.06}$ &...&...& $ 4.69^{+ 0.29}_{-0.27}$ & $359^{+ 61}_{-50}$ &$0.91( 99)$ & $ 1.15^{+ 0.06}_{-0.06}$  & $16^{+ 7}_{-5}$ \\
\hline
\multirow{3}{*}{C2} & MCD+PL& $ 0.15^{+ 0.14}_{-0.11}$ & $ 0.32^{+ 0.05}_{-0.04}$ & $ 2.22^{+ 3.32}_{-1.22}$ & $2.5$& $  9.9^{+ 10.3}_{ -9.9}$ &$0.98( 42)$ & $ 1.13^{+ 0.17}_{-0.13}$  & $ 1.84^{+ 0.67}_{-0.43}$ \\
&MCD& $ 0.09^{+ 0.11}_{-0.09}$ & $ 0.35^{+ 0.04}_{-0.04}$ & $ 1.43^{+ 1.34}_{-0.64}$ &...&...&$1.01( 43)$ & $ 1.14^{+ 0.19}_{-0.14}$  & $ 1.55^{+ 0.39}_{-0.27}$ \\
&PL& $ 0.72^{+ 0.18}_{-0.16}$ &...&...& $ 5.02^{+ 0.47}_{-0.40}$ & $471^{+185}_{-116}$ &$0.92( 43)$ & $ 1.02^{+ 0.11}_{-0.10}$  & $27^{+27}_{-12}$ \\
\cline{1-10}
\multirow{3}{*}{C3--C9} & MCD+PL & $ 0.14^{+ 0.03}_{-0.03}$ & $ 0.34^{+ 0.03}_{-0.02}$ & $ 0.80^{+ 0.34}_{-0.22}$ & $2.5$& $ 14.5^{+  3.7}_{ -3.6}$ &$0.97(137)$ & $ 0.64^{+ 0.02}_{-0.02}$  & $ 1.00^{+ 0.09}_{-0.08}$ \\
&MCD& $ 0.07^{+ 0.03}_{-0.02}$ & $ 0.40^{+ 0.02}_{-0.02}$ & $ 0.39^{+ 0.10}_{-0.08}$ &...&...&$1.25(138)$ & $ 0.60^{+ 0.02}_{-0.02}$  & $ 0.74^{+ 0.04}_{-0.04}$ \\
&PL& $ 0.47^{+ 0.04}_{-0.04}$ &...&...& $ 4.09^{+ 0.13}_{-0.13}$ & $147^{+ 13}_{-12}$ &$0.86(138)$ & $ 0.61^{+ 0.02}_{-0.02}$  & $ 4.57^{+ 0.77}_{-0.62}$ \\
\cline{1-10}
\multirow{3}{*}{C10} & MCD+PL & $ 0.46^{+ 0.27}_{-0.23}$ & $ 0.13^{+ 0.03}_{-0.02}$ & $461^{+4660}_{-411}$ & $2.5$& $  2.3^{+  1.8}_{ -1.3}$ &... & $ 0.31^{+ 0.06}_{-0.05}$  & $ 3.52^{+ 9.45}_{-2.31}$ \\
&MCD& $ 0.20^{+ 0.21}_{-0.18}$ & $ 0.17^{+ 0.04}_{-0.03}$ & $27^{+158}_{-22}$ &...&...&... & $ 0.32^{+ 0.10}_{-0.06}$  & $ 0.97^{+ 1.51}_{-0.53}$ \\
&PL& $ 0.81^{+ 0.32}_{-0.27}$ &...&...& $ 8.01^{+ 0.99}_{-1.42}$ & $136^{+134}_{-63}$ &... & $ 0.31^{+ 0.06}_{-0.05}$  & $96^{+688}_{-81}$ \\
\hline
\hline
\end{tabular}
\end{table*}

\clearpage

\begin{figure*}
\begin{center}
\includegraphics[width=6.3in]{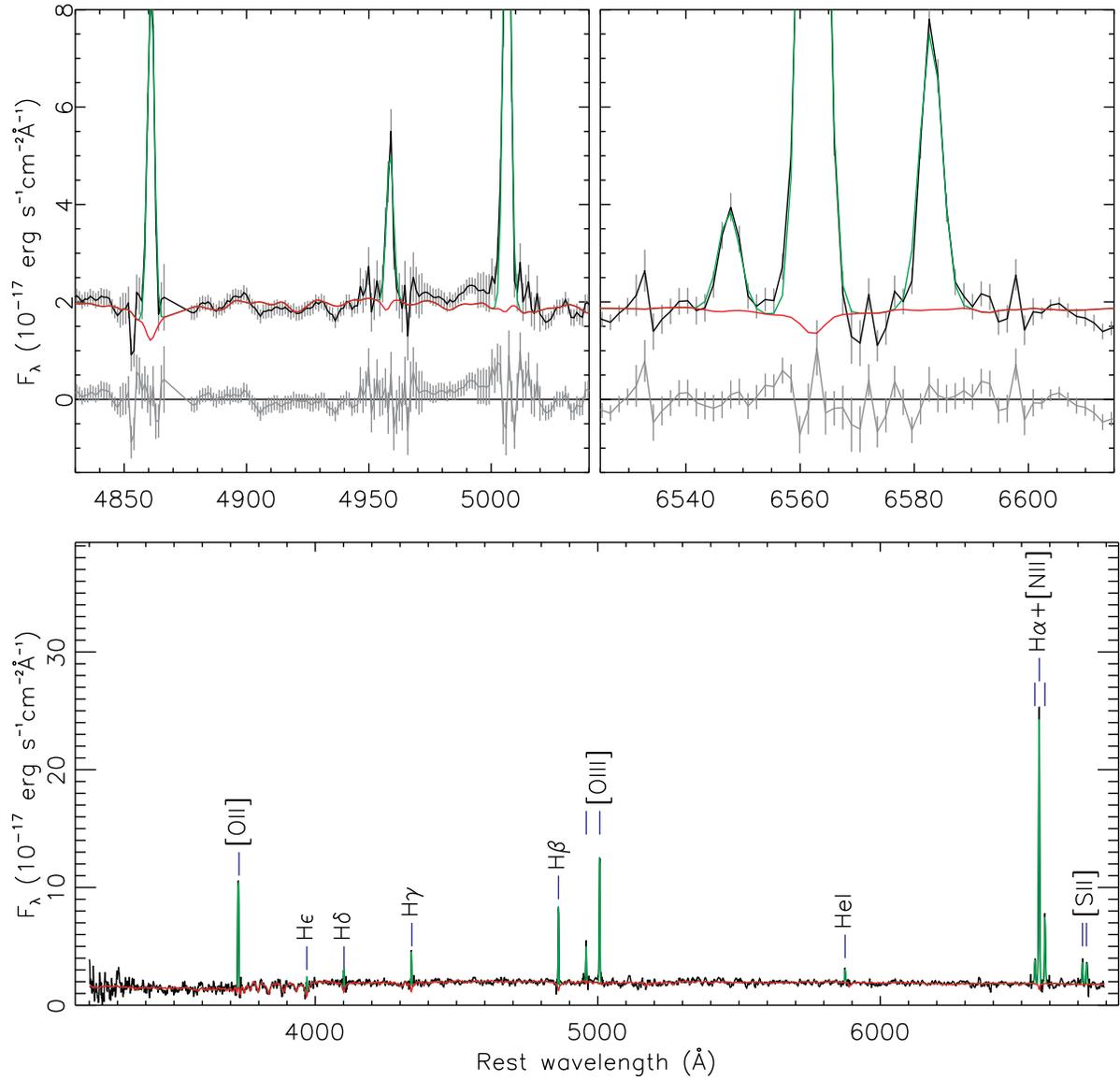}
\end{center}
\vskip -0.2in
\caption{\textbf{The SDSS optical spectrum of the candidate host galaxy of XJ1500+0154 taken on March 3rd 2011, showing only narrow emission lines consistent with a starburst galaxy.} The upper two panels zoom into the H$\beta$-[\ion{O}{III}] and H$\alpha$-[\ion{N}{II}] regions, with the fit residuals. The pPXF fit is shown as a solid green line, while the star component is shown as a red line. The data points outside the emission line regions have been smoothed with a box function of width 5, for clarity.
  \label{fig:s132740gsp}}
\end{figure*}

\clearpage
\begin{figure*}
\begin{center}
\includegraphics[width=6.3in]{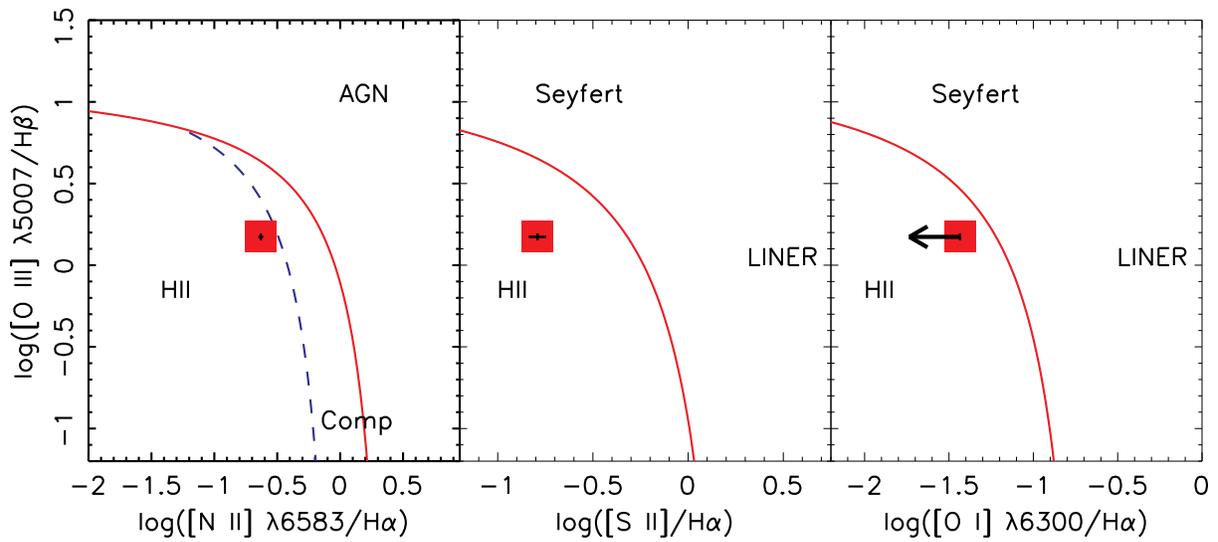}
\end{center}
\vskip -0.2in
\caption{\textbf{XJ1500+0154 on the BPT diagrams, indicating it as a
    star-forming galaxy.} The dashed and solid lines are used to
  separate galaxies into HII-region-like, AGN, and composite
  types\cite{kegrka2006} \label{fig:bpt}}
\end{figure*}

\clearpage
\begin{figure*}
\begin{center}
\includegraphics[width=6.3in]{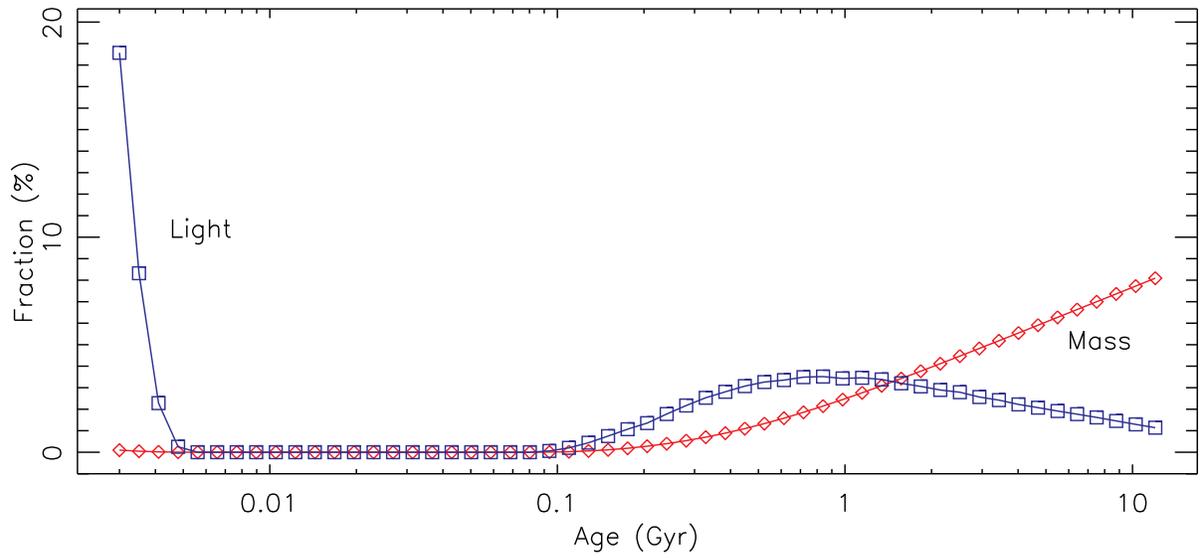}
\end{center}
\vskip -0.2in
\caption{\textbf{The distribution of the mass and light with respect to the age from the pPXF fit to the SDSS optical spectrum of the candidate host galaxy of XJ1500+0154, indicating the presence of very young ($<5$ Myr) populations.} The light was integrated over source rest-frame 3200 \AA\ and 6800 \AA. 
  \label{fig:s132740_agefraction}}
\end{figure*}

\clearpage

\begin{figure*}[tb]
\begin{center}
\includegraphics[width=5.6in]{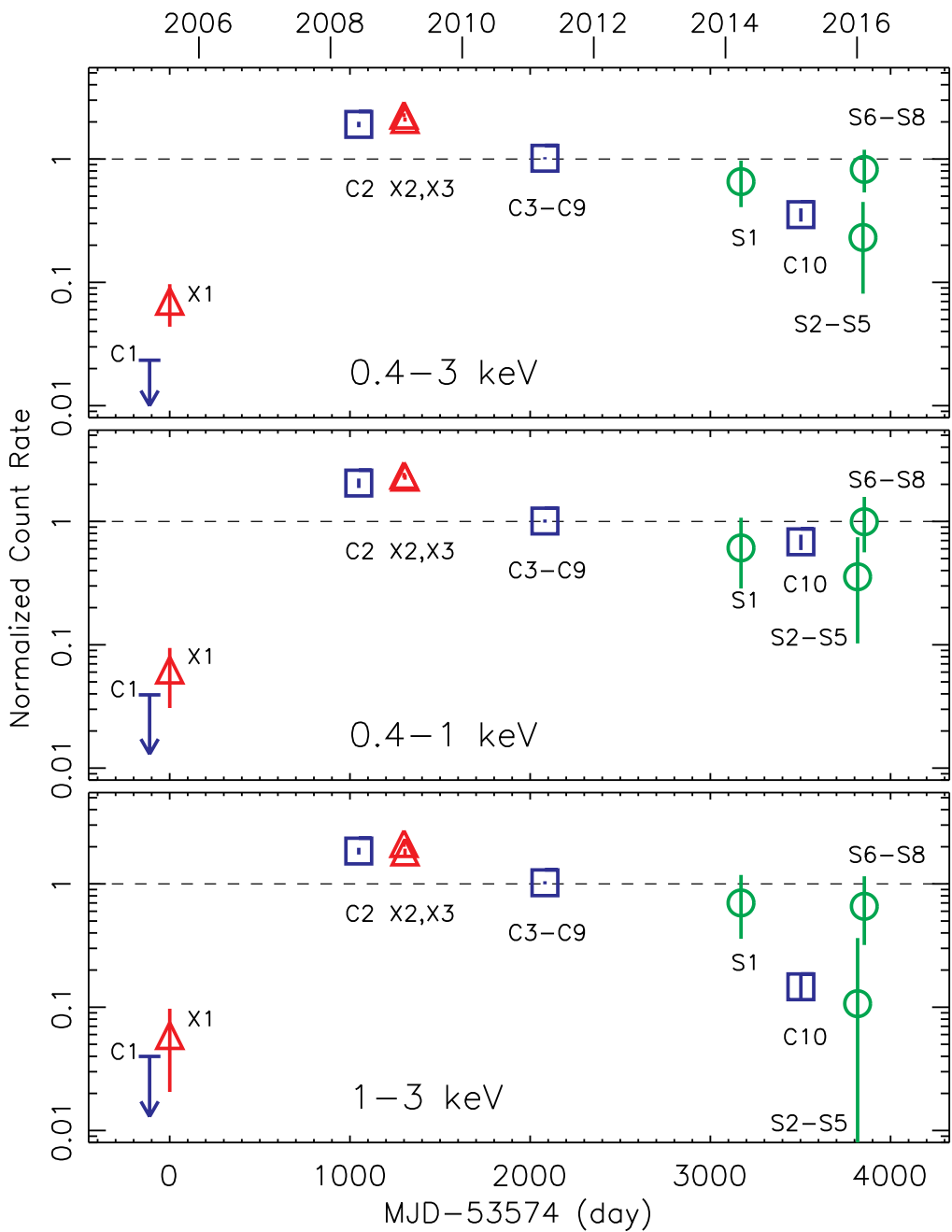}
\end{center}
\vskip -0.2in
\caption{\textbf{The long-term evolution of the count rates in 0.4--3 keV, 0.4--1 keV and
  1--3 keV (observer frame), normalized to those expected assuming the CompTT fit to
  C3--C9, with 90\% errors or $3\sigma$ upper limit (for C1).} For \emph{XMM-Newton} observations, the normalized count rates are the mean of all available cameras weighted by the error.  \label{fig:ltcr}}
\end{figure*}

\begin{figure*}[!htpb]
  \centering
  \subfigure[]{%
    \includegraphics[width=0.45\textwidth]{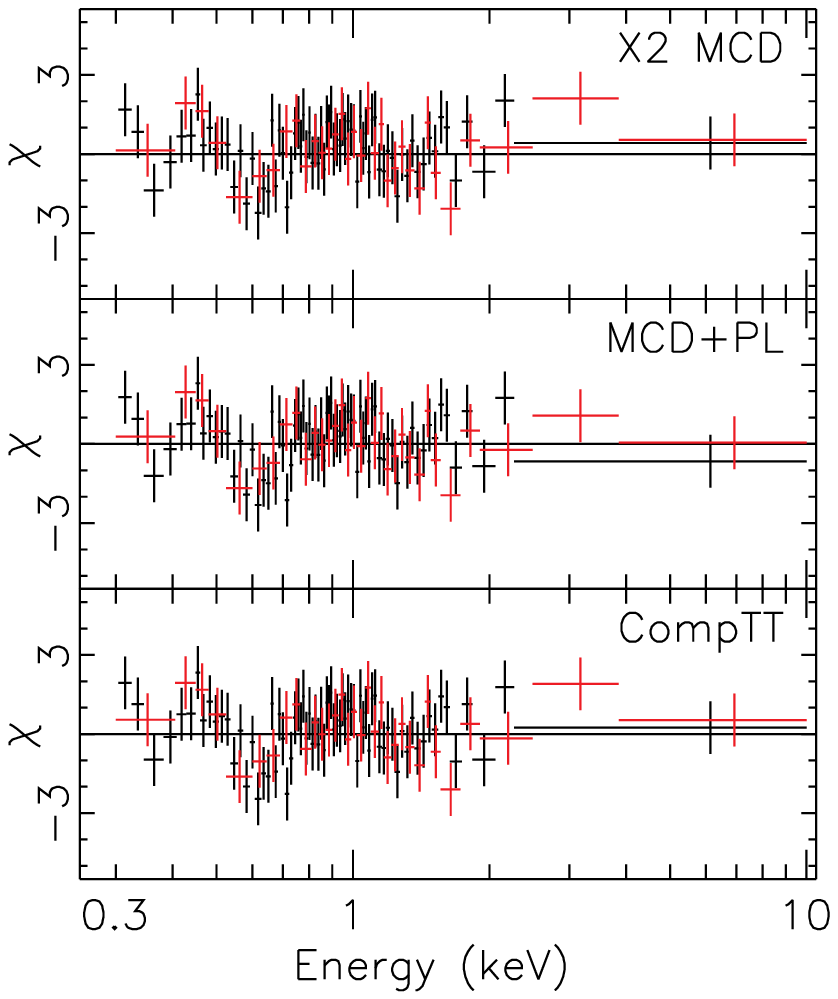}%
    \label{fig:delchix2}
  }
  \subfigure[]{%
    \includegraphics[width=0.45\textwidth]{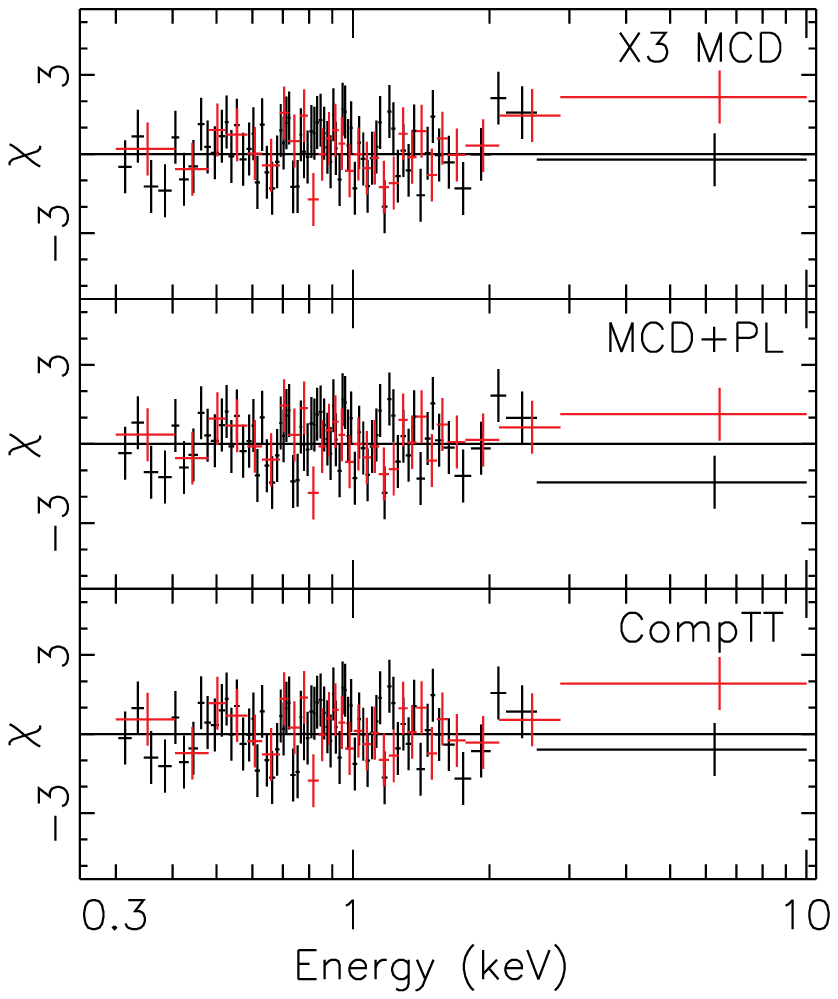}%
    \label{fig:delchix3}
  }\\
    \subfigure[]{%
    \includegraphics[width=0.45\textwidth]{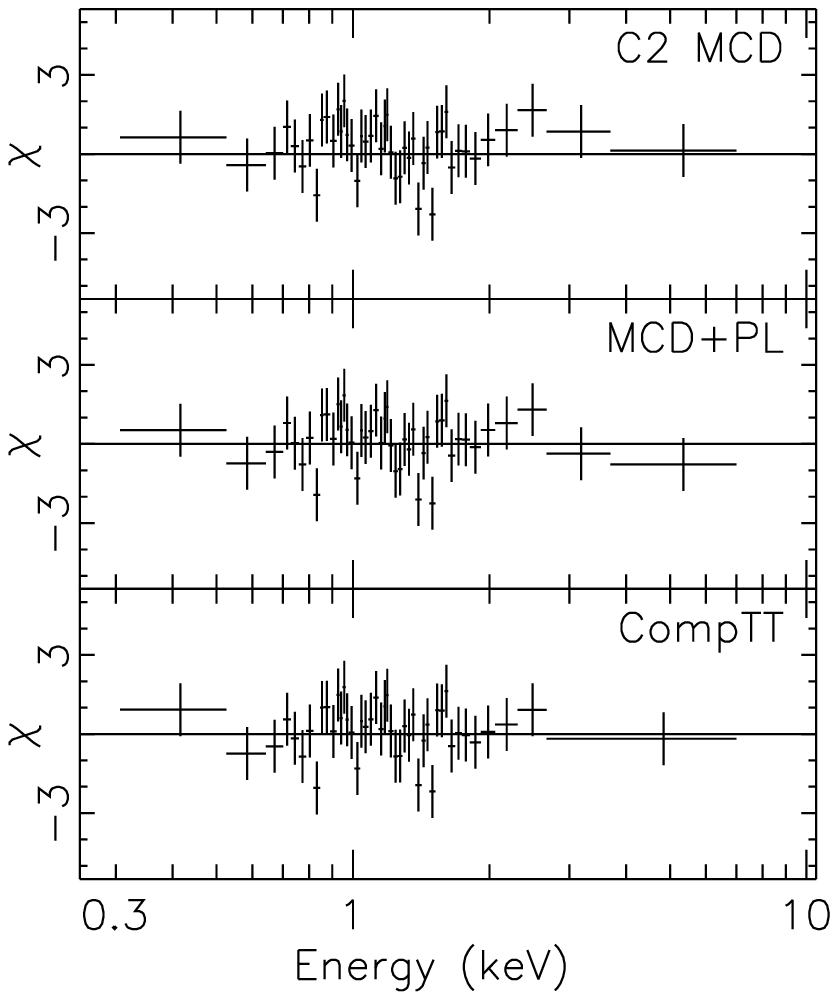}%
    \label{fig:delchic2}
  }
  \subfigure[]{%
    \includegraphics[width=0.45\textwidth]{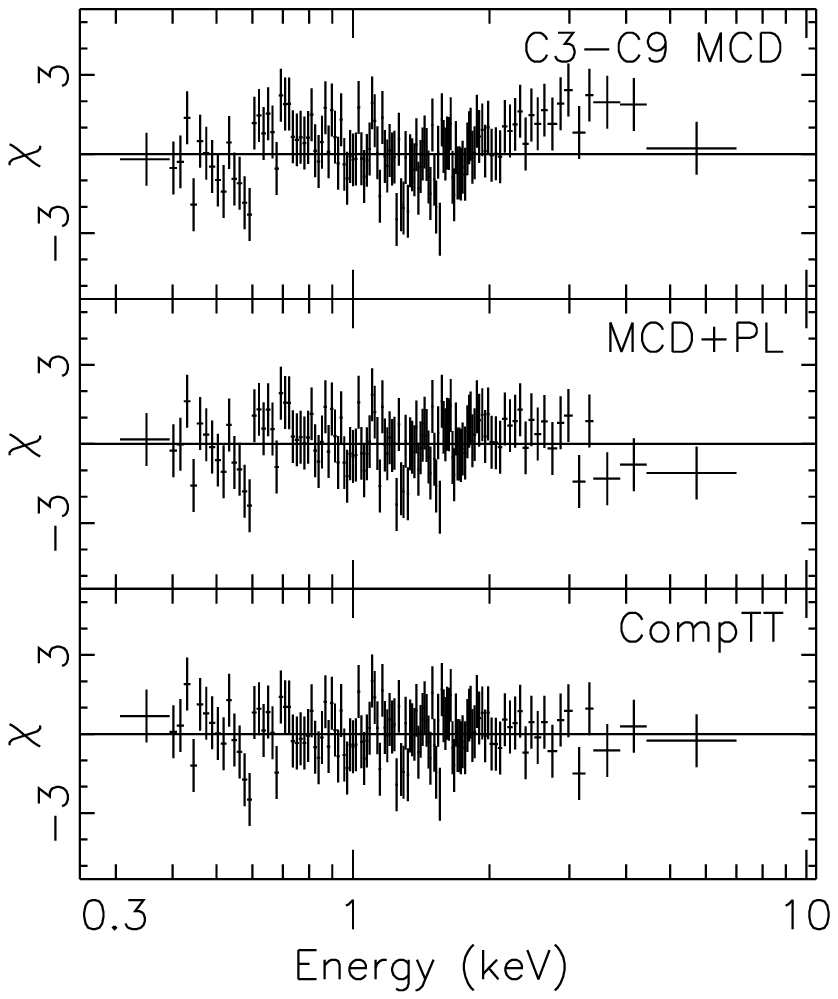}%
    \label{fig:delchic3c9}
  }
  \caption{\textbf{The fit residuals for the X2, X3, C2, and C3--C9 spectra, with the MCD, MCD+PL, and CompTT models.}\label{fig:delchi}}
\end{figure*}

\begin{figure*}
\begin{center}
\includegraphics[width=5.8in]{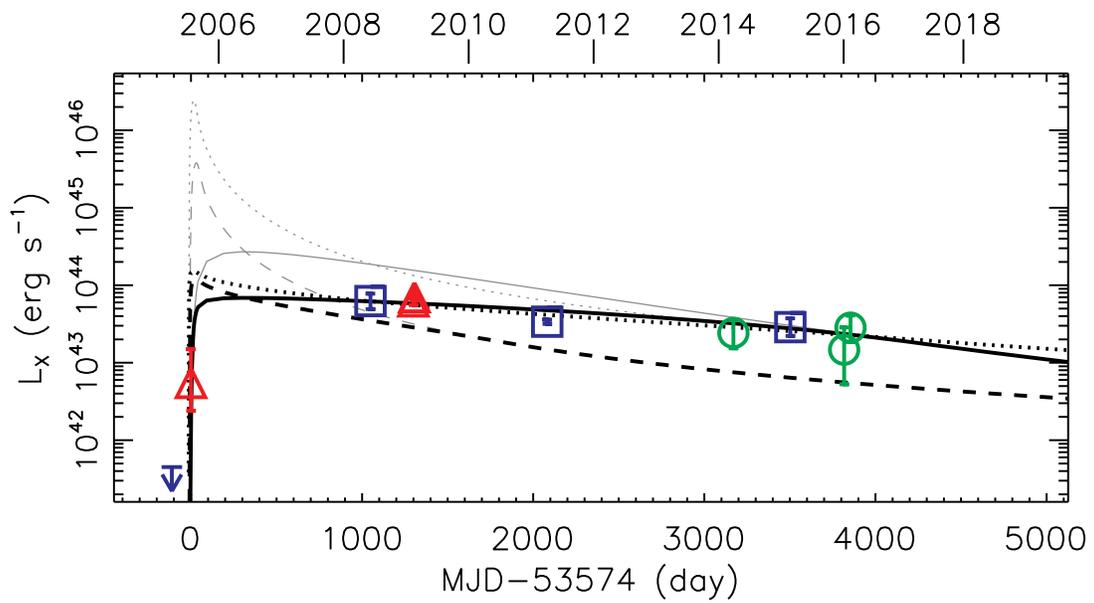}
\end{center}
\vskip -0.2in
\caption{\textbf{The modeling of the long-term X-ray luminosity curve
    of XJ1500+0154.}  The meanings of the symbols are the same as in
  Figure~\ref{fig:lumlcsp}.  The lines are for different models: solid
  lines for a 2 \msun\ star with slow circularization, dashed lines for a
  2 \msun\ star with prompt circularization, and dotted lines for a 10
  \msun\ star with prompt circularization. The thick black lines
  incorporate the super-Eddington accretion effects, while the thin
  gray lines do not (see \textit{SI}); they deviate from each other when the accretion rate is super-Eddington. All models assume a SMBH of mass $10^6$ \msun.
  \label{fig:lumlcmodel}  }
\end{figure*}

\clearpage
\begin{figure*}
\begin{center}
\includegraphics[width=6.8in]{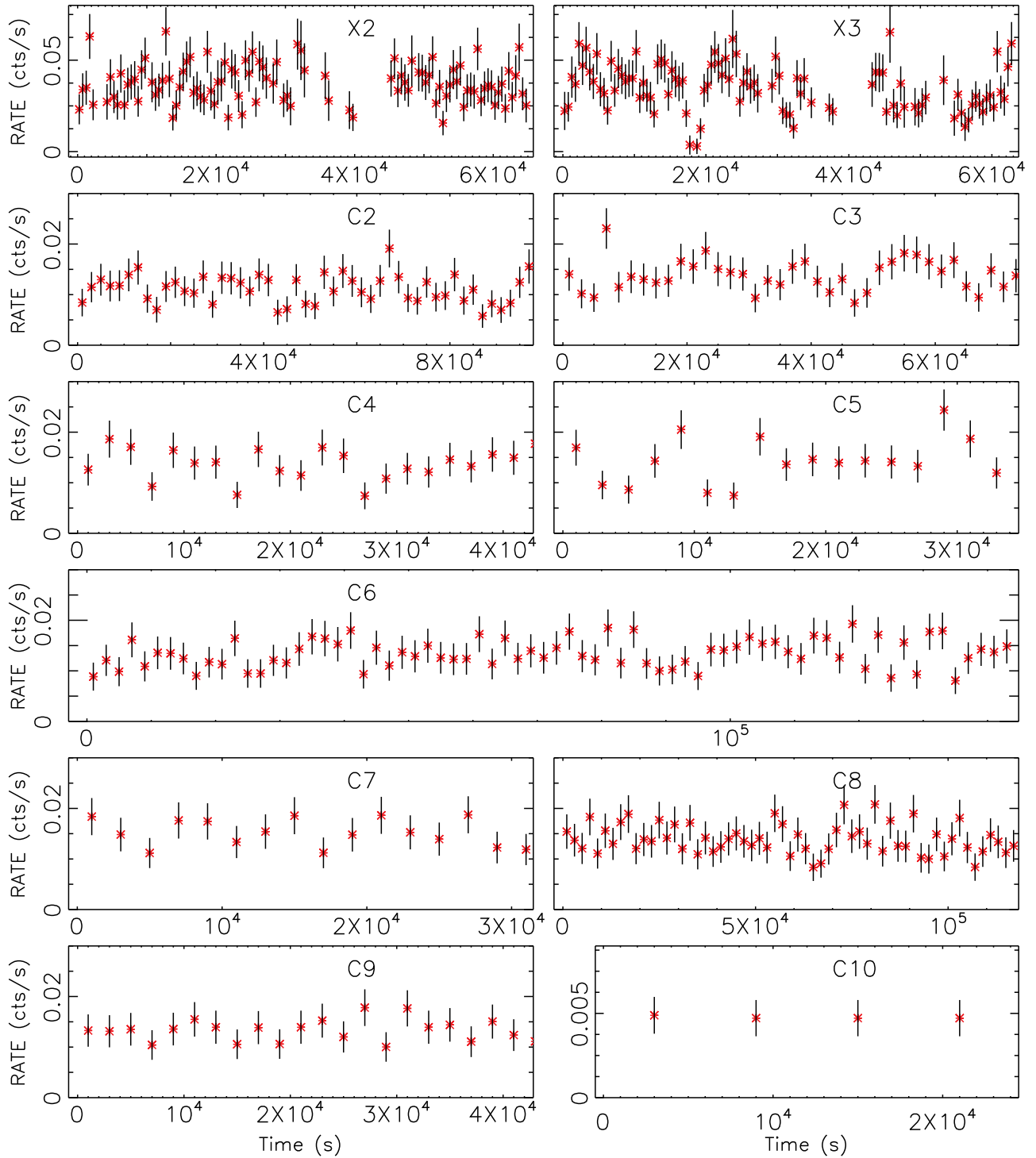}
\end{center}
\vskip -0.2in
\caption{\textbf{The background subtracted light curves from bright
    observations (0.3--3 keV for \emph{XMM-Newton} observations and
    0.4--3 keV for \emph{Chandra} observations).} For X2 and X3, we
  only show the pn data for clarity, and data in the strong background
  intervals have been excluded. The light curve bin size is 500 s for
  X2 and X3, 2 ks for C2--C9, and 6 ks for C10.  There seems to be a
  fast drop to zero count rate at 20 ks into the X3
  observation.  \label{fig:srclc}}
\end{figure*}

\clearpage
\begin{figure*}
\begin{center}
\includegraphics[width=5.5in]{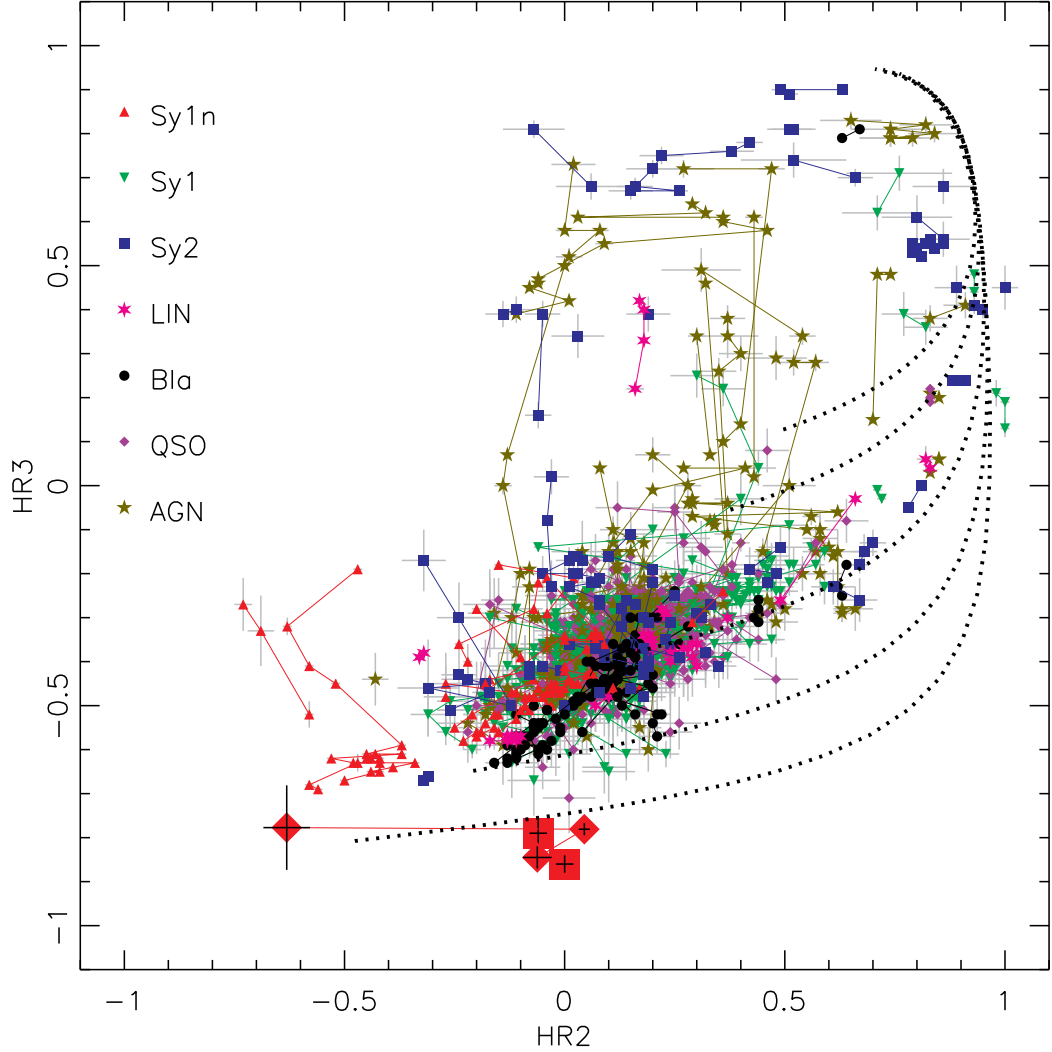}
\end{center}
\vskip -0.2in
\caption{\textbf{The X-ray color-color diagram for identified AGNs in
    Lin et al. 2012\cite{liweba2012} and XJ1500+0154 (big red squares)
    from the 3XMM-DR5 catalog.}  The colors HR2 and HR3 are defined
  using $(H-S)/(H+S)$, with $S$ and $H$ being the MOS1-medium-filter
  equivalent 0.5--1 keV and 1--2 keV counts rates for HR2 and 1--2 keV
  and 2--4.5 keV count rates for HR3, respectively. The \emph{Chandra}
  C2, C3--C9 and C10 observations are shown as big red diamonds. We
  overplot PL spectra (dotted lines) with $\Gamma_{\rm PL}=0.5$ (top),
  1, 2, 3, and 4 and $N_{\rm H}$ varying from 0 (lower-left) to
  $10^{23}$ cm$^{-2}$. The detections for each source are connected by
  solid lines in an increasing order of HR3. We only show AGN
  detections with S/N$\ge$18 (based on the 0.2--4.5 keV flux),
  resulting in 2002 detections in total. XJ1500+0154 has lower values
  of HR3 (thus softer in 1--4.5 keV) than AGNs in all observations
  shown.
  \label{fig:agnxraycolor}}
\end{figure*}


\begin{thebibliography}{100}
\expandafter\ifx\csname url\endcsname\relax
  \def\url#1{\texttt{#1}}\fi
\expandafter\ifx\csname urlprefix\endcsname\relax\def\urlprefix{URL }\fi
\providecommand{\bibinfo}[2]{#2}
\providecommand{\eprint}[2][]{\url{#2}}

\bibitem{re1988}
\bibinfo{author}{{Rees}, M.~J.}
\newblock \bibinfo{title}{{Tidal disruption of stars by black holes of 10 to
  the 6th-10 to the 8th solar masses in nearby galaxies}}.
\newblock \emph{\bibinfo{journal}{\nat}} \textbf{\bibinfo{volume}{333}},
  \bibinfo{pages}{523--528} (\bibinfo{year}{1988}).

\bibitem{gechre2012}
\bibinfo{author}{{Gezari}, S.} \emph{et~al.}
\newblock \bibinfo{title}{{An ultraviolet-optical flare from the tidal
  disruption of a helium-rich stellar core}}.
\newblock \emph{\bibinfo{journal}{\nat}} \textbf{\bibinfo{volume}{485}},
  \bibinfo{pages}{217--220} (\bibinfo{year}{2012}).
\newblock \eprint{1205.0252}.

\bibitem{zabema2013}
\bibinfo{author}{{Zauderer}, B.~A.} \emph{et~al.}
\newblock \bibinfo{title}{{Radio Monitoring of the Tidal Disruption Event Swift
  J164449.3+573451. II. The Relativistic Jet Shuts Off and a Transition to
  Forward Shock X-Ray/Radio Emission}}.
\newblock \emph{\bibinfo{journal}{\apj}} \textbf{\bibinfo{volume}{767}},
  \bibinfo{pages}{152} (\bibinfo{year}{2013}).
\newblock \eprint{1212.1173}.

\bibitem{mikami2015}
\bibinfo{author}{{Miller}, J.~M.} \emph{et~al.}
\newblock \bibinfo{title}{{Flows of X-ray gas reveal the disruption of a star
  by a massive black hole}}.
\newblock \emph{\bibinfo{journal}{\nat}} \textbf{\bibinfo{volume}{526}},
  \bibinfo{pages}{542--545} (\bibinfo{year}{2015}).
\newblock \eprint{1510.06348}.

\bibitem{vaanst2016}
\bibinfo{author}{{van Velzen}, S.} \emph{et~al.}
\newblock \bibinfo{title}{{A radio jet from the optical and x-ray bright
  stellar tidal disruption flare ASASSN-14li}}.
\newblock \emph{\bibinfo{journal}{Science}} \textbf{\bibinfo{volume}{351}},
  \bibinfo{pages}{62--65} (\bibinfo{year}{2016}).
\newblock \eprint{1511.08803}.

\bibitem{mimima2013}
\bibinfo{author}{{Middleton}, M.~J.} \emph{et~al.}
\newblock \bibinfo{title}{{Bright radio emission from an ultraluminous
  stellar-mass microquasar in M 31}}.
\newblock \emph{\bibinfo{journal}{\nat}} \textbf{\bibinfo{volume}{493}},
  \bibinfo{pages}{187--190} (\bibinfo{year}{2013}).
\newblock \eprint{1212.4698}.

\bibitem{ranujo2015}
\bibinfo{author}{{Randall}, S.~W.} \emph{et~al.}
\newblock \bibinfo{title}{{A Very Deep Chandra Observation of the Galaxy Group
  NGC 5813: AGN Shocks, Feedback, and Outburst History}}.
\newblock \emph{\bibinfo{journal}{\apj}} \textbf{\bibinfo{volume}{805}},
  \bibinfo{pages}{112} (\bibinfo{year}{2015}).
\newblock \eprint{1503.08205}.

\bibitem{revo2015}
\bibinfo{author}{{Reines}, A.~E.} \& \bibinfo{author}{{Volonteri}, M.}
\newblock \bibinfo{title}{{Relations between Central Black Hole Mass and Total
  Galaxy Stellar Mass in the Local Universe}}.
\newblock \emph{\bibinfo{journal}{\apj}} \textbf{\bibinfo{volume}{813}},
  \bibinfo{pages}{82} (\bibinfo{year}{2015}).
\newblock \eprint{1508.06274}.

\bibitem{gido2004}
\bibinfo{author}{{Gierli{\'n}ski}, M.} \& \bibinfo{author}{{Done}, C.}
\newblock \bibinfo{title}{{Is the soft excess in active galactic nuclei real?}}
\newblock \emph{\bibinfo{journal}{\mnras}} \textbf{\bibinfo{volume}{349}},
  \bibinfo{pages}{L7--L11} (\bibinfo{year}{2004}).
\newblock \eprint{arXiv:astro-ph/0312271}.

\bibitem{ti1994}
\bibinfo{author}{{Titarchuk}, L.}
\newblock \bibinfo{title}{{Generalized Comptonization models and application to
  the recent high-energy observations}}.
\newblock \emph{\bibinfo{journal}{\apj}} \textbf{\bibinfo{volume}{434}},
  \bibinfo{pages}{570--586} (\bibinfo{year}{1994}).

\bibitem{glrodo2009}
\bibinfo{author}{{Gladstone}, J.~C.}, \bibinfo{author}{{Roberts}, T.~P.} \&
  \bibinfo{author}{{Done}, C.}
\newblock \bibinfo{title}{{The ultraluminous state}}.
\newblock \emph{\bibinfo{journal}{\mnras}} \textbf{\bibinfo{volume}{397}},
  \bibinfo{pages}{1836--1851} (\bibinfo{year}{2009}).
\newblock \eprint{0905.4076}.

\bibitem{liirwe2013}
\bibinfo{author}{{Lin}, D.}, \bibinfo{author}{{Irwin}, J.~A.},
  \bibinfo{author}{{Webb}, N.~A.}, \bibinfo{author}{{Barret}, D.} \&
  \bibinfo{author}{{Remillard}, R.~A.}
\newblock \bibinfo{title}{{Discovery of a Highly Variable Dipping Ultraluminous
  X-Ray Source in M94}}.
\newblock \emph{\bibinfo{journal}{\apj}} \textbf{\bibinfo{volume}{779}},
  \bibinfo{pages}{149} (\bibinfo{year}{2013}).
\newblock \eprint{1311.1198}.

\bibitem{kimu2016}
\bibinfo{author}{{King}, A.} \& \bibinfo{author}{{Muldrew}, S.~I.}
\newblock \bibinfo{title}{{Black hole winds II: Hyper-Eddington winds and
  feedback}}.
\newblock \emph{\bibinfo{journal}{\mnras}} \textbf{\bibinfo{volume}{455}},
  \bibinfo{pages}{1211--1217} (\bibinfo{year}{2016}).
\newblock \eprint{1510.01736}.

\bibitem{pimifa2016}
\bibinfo{author}{{Pinto}, C.}, \bibinfo{author}{{Middleton}, M.~J.} \&
  \bibinfo{author}{{Fabian}, A.~C.}
\newblock \bibinfo{title}{{Resolved atomic lines reveal outflows in two
  ultraluminous X-ray sources}}.
\newblock \emph{\bibinfo{journal}{\nat}}  (\bibinfo{year}{2016}).
\newblock \eprint{1604.08593}.

\bibitem{ul1999}
\bibinfo{author}{{Ulmer}, A.}
\newblock \bibinfo{title}{{Flares from the Tidal Disruption of Stars by Massive
  Black Holes}}.
\newblock \emph{\bibinfo{journal}{\apj}} \textbf{\bibinfo{volume}{514}},
  \bibinfo{pages}{180--187} (\bibinfo{year}{1999}).

\bibitem{ohmi2007}
\bibinfo{author}{{Ohsuga}, K.} \& \bibinfo{author}{{Mineshige}, S.}
\newblock \bibinfo{title}{{Why Is Supercritical Disk Accretion Feasible?}}
\newblock \emph{\bibinfo{journal}{\apj}} \textbf{\bibinfo{volume}{670}},
  \bibinfo{pages}{1283--1290} (\bibinfo{year}{2007}).
\newblock \eprint{arXiv:0710.2941}.

\bibitem{krpi2012}
\bibinfo{author}{{Krolik}, J.~H.} \& \bibinfo{author}{{Piran}, T.}
\newblock \bibinfo{title}{{Jets from Tidal Disruptions of Stars by Black
  Holes}}.
\newblock \emph{\bibinfo{journal}{\apj}} \textbf{\bibinfo{volume}{749}},
  \bibinfo{pages}{92} (\bibinfo{year}{2012}).
\newblock \eprint{1111.2802}.

\bibitem{ko1994}
\bibinfo{author}{{Kochanek}, C.~S.}
\newblock \bibinfo{title}{{The aftermath of tidal disruption: The dynamics of
  thin gas streams}}.
\newblock \emph{\bibinfo{journal}{\apj}} \textbf{\bibinfo{volume}{422}},
  \bibinfo{pages}{508--520} (\bibinfo{year}{1994}).

\bibitem{gura2015}
\bibinfo{author}{{Guillochon}, J.} \& \bibinfo{author}{{Ramirez-Ruiz}, E.}
\newblock \bibinfo{title}{{A Dark Year for Tidal Disruption Events}}.
\newblock \emph{\bibinfo{journal}{\apj}} \textbf{\bibinfo{volume}{809}},
  \bibinfo{pages}{166} (\bibinfo{year}{2015}).
\newblock \eprint{1501.05306}.

\bibitem{pisvkr2015}
\bibinfo{author}{{Piran}, T.}, \bibinfo{author}{{Svirski}, G.},
  \bibinfo{author}{{Krolik}, J.}, \bibinfo{author}{{Cheng}, R.~M.} \&
  \bibinfo{author}{{Shiokawa}, H.}
\newblock \bibinfo{title}{{Disk Formation Versus Disk Accretion: What Powers
  Tidal Disruption Events?}}
\newblock \emph{\bibinfo{journal}{\apj}} \textbf{\bibinfo{volume}{806}},
  \bibinfo{pages}{164} (\bibinfo{year}{2015}).
\newblock \eprint{1502.05792}.

\bibitem{shkrch2015}
\bibinfo{author}{{Shiokawa}, H.}, \bibinfo{author}{{Krolik}, J.~H.},
  \bibinfo{author}{{Cheng}, R.~M.}, \bibinfo{author}{{Piran}, T.} \&
  \bibinfo{author}{{Noble}, S.~C.}
\newblock \bibinfo{title}{{General Relativistic Hydrodynamic Simulation of
  Accretion Flow from a Stellar Tidal Disruption}}.
\newblock \emph{\bibinfo{journal}{\apj}} \textbf{\bibinfo{volume}{804}},
  \bibinfo{pages}{85} (\bibinfo{year}{2015}).
\newblock \eprint{1501.04365}.

\bibitem{hastlo2016}
\bibinfo{author}{{Hayasaki}, K.}, \bibinfo{author}{{Stone}, N.} \&
  \bibinfo{author}{{Loeb}, A.}
\newblock \bibinfo{title}{{Circularization of tidally disrupted stars around
  spinning supermassive black holes}}.
\newblock \emph{\bibinfo{journal}{\mnras}} \textbf{\bibinfo{volume}{461}},
  \bibinfo{pages}{3760--3780} (\bibinfo{year}{2016}).
\newblock \eprint{1501.05207}.

\bibitem{liname2002}
\bibinfo{author}{{Li}, L.-X.}, \bibinfo{author}{{Narayan}, R.} \&
  \bibinfo{author}{{Menou}, K.}
\newblock \bibinfo{title}{{The Giant X-Ray Flare of NGC 5905: Tidal Disruption
  of a Star, a Brown Dwarf, or a Planet?}}
\newblock \emph{\bibinfo{journal}{\apj}} \textbf{\bibinfo{volume}{576}},
  \bibinfo{pages}{753--761} (\bibinfo{year}{2002}).
\newblock \eprint{astro-ph/0203191}.

\bibitem{kohasc2004}
\bibinfo{author}{{Komossa}, S.} \emph{et~al.}
\newblock \bibinfo{title}{{A Huge Drop in the X-Ray Luminosity of the Nonactive
  Galaxy RX J1242.6-1119A, and the First Postflare Spectrum: Testing the Tidal
  Disruption Scenario}}.
\newblock \emph{\bibinfo{journal}{\apjl}} \textbf{\bibinfo{volume}{603}},
  \bibinfo{pages}{L17--L20} (\bibinfo{year}{2004}).
\newblock \eprint{arXiv:astro-ph/0402468}.

\bibitem{dobrer2002}
\bibinfo{author}{{Donley}, J.~L.}, \bibinfo{author}{{Brandt}, W.~N.},
  \bibinfo{author}{{Eracleous}, M.} \& \bibinfo{author}{{Boller}, T.}
\newblock \bibinfo{title}{{Large-Amplitude X-Ray Outbursts from Galactic
  Nuclei: A Systematic Survey using ROSAT Archival Data}}.
\newblock \emph{\bibinfo{journal}{\aj}} \textbf{\bibinfo{volume}{124}},
  \bibinfo{pages}{1308--1321} (\bibinfo{year}{2002}).
\newblock \eprint{astro-ph/0206291}.

\bibitem{ko2016}
\bibinfo{author}{{Kochanek}, C.~S.}
\newblock \bibinfo{title}{{Tidal disruption event demographics}}.
\newblock \emph{\bibinfo{journal}{\mnras}} \textbf{\bibinfo{volume}{461}},
  \bibinfo{pages}{371--384} (\bibinfo{year}{2016}).
\newblock \eprint{1601.06787}.

\bibitem{mowave2011}
\bibinfo{author}{{Mortlock}, D.~J.} \emph{et~al.}
\newblock \bibinfo{title}{{A luminous quasar at a redshift of z = 7.085}}.
\newblock \emph{\bibinfo{journal}{\nat}} \textbf{\bibinfo{volume}{474}},
  \bibinfo{pages}{616--619} (\bibinfo{year}{2011}).
\newblock \eprint{1106.6088}.

\bibitem{vore2005}
\bibinfo{author}{{Volonteri}, M.} \& \bibinfo{author}{{Rees}, M.~J.}
\newblock \bibinfo{title}{{Rapid Growth of High-Redshift Black Holes}}.
\newblock \emph{\bibinfo{journal}{\apj}} \textbf{\bibinfo{volume}{633}},
  \bibinfo{pages}{624--629} (\bibinfo{year}{2005}).
\newblock \eprint{astro-ph/0506040}.

\bibitem{marala2005}
\bibinfo{author}{{Mart{\'{\i}}nez-Sansigre}, A.} \emph{et~al.}
\newblock \bibinfo{title}{{The obscuration by dust of most of the growth of
  supermassive black holes}}.
\newblock \emph{\bibinfo{journal}{\nat}} \textbf{\bibinfo{volume}{436}},
  \bibinfo{pages}{666--669} (\bibinfo{year}{2005}).
\newblock \eprint{astro-ph/0505486}.

\bibitem{mauler2013}
\bibinfo{author}{{Maksym}, W.~P.}, \bibinfo{author}{{Ulmer}, M.~P.},
  \bibinfo{author}{{Eracleous}, M.~C.}, \bibinfo{author}{{Guennou}, L.} \&
  \bibinfo{author}{{Ho}, L.~C.}
\newblock \bibinfo{title}{{A tidal flare candidate in Abell 1795}}.
\newblock \emph{\bibinfo{journal}{\mnras}} \textbf{\bibinfo{volume}{435}},
  \bibinfo{pages}{1904--1927} (\bibinfo{year}{2013}).
\newblock \eprint{1307.6556}.
\end{thebibliography}

\begin{thebibliography}{10}
\expandafter\ifx\csname url\endcsname\relax
  \def\url#1{\texttt{#1}}\fi
\expandafter\ifx\csname urlprefix\endcsname\relax\def\urlprefix{URL }\fi
\providecommand{\bibinfo}[2]{#2}
\providecommand{\eprint}[2][]{\url{#2}}
\setcounter{enumiv}{30}

\bibitem{jalual2001}
\bibinfo{author}{{Jansen}, F.} \emph{et~al.}
\newblock \bibinfo{title}{{XMM-Newton observatory. I. The spacecraft and
  operations}}.
\newblock \emph{\bibinfo{journal}{\aap}} \textbf{\bibinfo{volume}{365}},
  \bibinfo{pages}{L1--L6} (\bibinfo{year}{2001}).

\bibitem{stbrde2001}
\bibinfo{author}{{Str{\"u}der}, L.} \emph{et~al.}
\newblock \bibinfo{title}{{The European Photon Imaging Camera on XMM-Newton:
  The pn-CCD camera}}.
\newblock \emph{\bibinfo{journal}{\aap}} \textbf{\bibinfo{volume}{365}},
  \bibinfo{pages}{L18--L26} (\bibinfo{year}{2001}).

\bibitem{tuabar2001}
\bibinfo{author}{{Turner}, M.~J.~L.} \emph{et~al.}
\newblock \bibinfo{title}{{The European Photon Imaging Camera on XMM-Newton:
  The MOS cameras : The MOS cameras}}.
\newblock \emph{\bibinfo{journal}{\aap}} \textbf{\bibinfo{volume}{365}},
  \bibinfo{pages}{L27--L35} (\bibinfo{year}{2001}).
\newblock \eprint{arXiv:astro-ph/0011498}.

\bibitem{wascfy2009}
\bibinfo{author}{{Watson}, M.~G.} \emph{et~al.}
\newblock \bibinfo{title}{{The XMM-Newton serendipitous survey. V. The Second
  XMM-Newton serendipitous source catalogue}}.
\newblock \emph{\bibinfo{journal}{\aap}} \textbf{\bibinfo{volume}{493}},
  \bibinfo{pages}{339--373} (\bibinfo{year}{2009}).
\newblock \eprint{0807.1067}.

\bibitem{bapiba1998}
\bibinfo{author}{{Bautz}, M.~W.} \emph{et~al.}
\newblock \bibinfo{title}{{X-ray CCD calibration for the AXAF CCD Imaging
  Spectrometer}}.
\newblock In \bibinfo{editor}{{R.~B.~Hoover \& A.~B.~Walker}} (ed.)
  \emph{\bibinfo{booktitle}{Society of Photo-Optical Instrumentation Engineers
  (SPIE) Conference Series}}, vol. \bibinfo{volume}{3444} of
  \emph{\bibinfo{series}{Society of Photo-Optical Instrumentation Engineers
  (SPIE) Conference Series}}, \bibinfo{pages}{210--224} (\bibinfo{year}{1998}).

\bibitem{grlo1992}
\bibinfo{author}{{Gregory}, P.~C.} \& \bibinfo{author}{{Loredo}, T.~J.}
\newblock \bibinfo{title}{{A new method for the detection of a periodic signal
  of unknown shape and period}}.
\newblock \emph{\bibinfo{journal}{\apj}} \textbf{\bibinfo{volume}{398}},
  \bibinfo{pages}{146--168} (\bibinfo{year}{1992}).

\bibitem{evprgl2010}
\bibinfo{author}{{Evans}, I.~N.} \emph{et~al.}
\newblock \bibinfo{title}{{The Chandra Source Catalog}}.
\newblock \emph{\bibinfo{journal}{\apjs}} \textbf{\bibinfo{volume}{189}},
  \bibinfo{pages}{37--82} (\bibinfo{year}{2010}).
\newblock \eprint{1005.4665}.

\bibitem{frkaro2002}
\bibinfo{author}{{Freeman}, P.~E.}, \bibinfo{author}{{Kashyap}, V.},
  \bibinfo{author}{{Rosner}, R.} \& \bibinfo{author}{{Lamb}, D.~Q.}
\newblock \bibinfo{title}{{A Wavelet-Based Algorithm for the Spatial Analysis
  of Poisson Data}}.
\newblock \emph{\bibinfo{journal}{\apjs}} \textbf{\bibinfo{volume}{138}},
  \bibinfo{pages}{185--218} (\bibinfo{year}{2002}).
\newblock \eprint{arXiv:astro-ph/0108429}.

\bibitem{bochab2003}
\bibinfo{author}{{Boulade}, O.} \emph{et~al.}
\newblock \bibinfo{title}{{MegaCam: the new Canada-France-Hawaii Telescope
  wide-field imaging camera}}.
\newblock In \bibinfo{editor}{{Iye}, M.} \& \bibinfo{editor}{{Moorwood},
  A.~F.~M.} (eds.) \emph{\bibinfo{booktitle}{Instrument Design and Performance
  for Optical/Infrared Ground-based Telescopes}}, vol. \bibinfo{volume}{4841}
  of \emph{\bibinfo{series}{Society of Photo-Optical Instrumentation Engineers
  (SPIE) Conference Series}}, \bibinfo{pages}{72--81} (\bibinfo{year}{2003}).

\bibitem{rafogi2011}
\bibinfo{author}{{Randall}, S.~W.} \emph{et~al.}
\newblock \bibinfo{title}{{Shocks and Cavities from Multiple Outbursts in the
  Galaxy Group NGC 5813: A Window to Active Galactic Nucleus Feedback}}.
\newblock \emph{\bibinfo{journal}{\apj}} \textbf{\bibinfo{volume}{726}},
  \bibinfo{pages}{86} (\bibinfo{year}{2011}).
\newblock \eprint{1006.4379}.

\bibitem{kikiwi2007}
\bibinfo{author}{{Kim}, M.} \emph{et~al.}
\newblock \bibinfo{title}{{Chandra Multiwavelength Project X-Ray Point Source
  Catalog}}.
\newblock \emph{\bibinfo{journal}{\apjs}} \textbf{\bibinfo{volume}{169}},
  \bibinfo{pages}{401--429} (\bibinfo{year}{2007}).

\bibitem{licawe2016}
\bibinfo{author}{{Lin}, D.} \emph{et~al.}
\newblock \bibinfo{title}{{Discovery of the Candidate Off-nuclear Ultrasoft
  Hyper-luminous X-ray Source 3XMM J141711.1+522541}}.
\newblock \emph{\bibinfo{journal}{\apj}} \textbf{\bibinfo{volume}{821}},
  \bibinfo{pages}{25} (\bibinfo{year}{2016}).
\newblock \eprint{1603.00455}.

\bibitem{gechgi2004}
\bibinfo{author}{{Gehrels}, N.} \emph{et~al.}
\newblock \bibinfo{title}{{The Swift Gamma-Ray Burst Mission}}.
\newblock \emph{\bibinfo{journal}{\apj}} \textbf{\bibinfo{volume}{611}},
  \bibinfo{pages}{1005--1020} (\bibinfo{year}{2004}).

\bibitem{buhino2005}
\bibinfo{author}{{Burrows}, D.~N.} \emph{et~al.}
\newblock \bibinfo{title}{{The Swift X-Ray Telescope}}.
\newblock \emph{\bibinfo{journal}{\ssr}} \textbf{\bibinfo{volume}{120}},
  \bibinfo{pages}{165--195} (\bibinfo{year}{2005}).
\newblock \eprint{arXiv:astro-ph/0508071}.

\bibitem{rokema2005}
\bibinfo{author}{{Roming}, P.~W.~A.} \emph{et~al.}
\newblock \bibinfo{title}{{The Swift Ultra-Violet/Optical Telescope}}.
\newblock \emph{\bibinfo{journal}{\ssr}} \textbf{\bibinfo{volume}{120}},
  \bibinfo{pages}{95--142} (\bibinfo{year}{2005}).
\newblock \eprint{arXiv:astro-ph/0507413}.

\bibitem{gw2008}
\bibinfo{author}{{Gwyn}, S.~D.~J.}
\newblock \bibinfo{title}{{MegaPipe: The MegaCam Image Stacking Pipeline at the
  Canadian Astronomical Data Centre}}.
\newblock \emph{\bibinfo{journal}{\pasp}} \textbf{\bibinfo{volume}{120}},
  \bibinfo{pages}{212--223} (\bibinfo{year}{2008}).
\newblock \eprint{0710.0370}.

\bibitem{abadag2009}
\bibinfo{author}{{Abazajian}, K.~N.} \emph{et~al.}
\newblock \bibinfo{title}{{The Seventh Data Release of the Sloan Digital Sky
  Survey}}.
\newblock \emph{\bibinfo{journal}{\apjs}} \textbf{\bibinfo{volume}{182}},
  \bibinfo{pages}{543--558} (\bibinfo{year}{2009}).
\newblock \eprint{0812.0649}.

\bibitem{pehoim2010}
\bibinfo{author}{{Peng}, C.~Y.}, \bibinfo{author}{{Ho}, L.~C.},
  \bibinfo{author}{{Impey}, C.~D.} \& \bibinfo{author}{{Rix}, H.-W.}
\newblock \bibinfo{title}{{Detailed Decomposition of Galaxy Images. II. Beyond
  Axisymmetric Models}}.
\newblock \emph{\bibinfo{journal}{\aj}} \textbf{\bibinfo{volume}{139}},
  \bibinfo{pages}{2097--2129} (\bibinfo{year}{2010}).
\newblock \eprint{0912.0731}.

\bibitem{ar1996}
\bibinfo{author}{{Arnaud}, K.~A.}
\newblock \bibinfo{title}{{XSPEC: The First Ten Years}}.
\newblock In \bibinfo{editor}{{Jacoby}, G.~H.} \& \bibinfo{editor}{{Barnes},
  J.} (eds.) \emph{\bibinfo{booktitle}{Astronomical Data Analysis Software and
  Systems V}}, vol. \bibinfo{volume}{101} of
  \emph{\bibinfo{series}{Astronomical Society of the Pacific Conference
  Series}}, \bibinfo{pages}{17--+} (\bibinfo{year}{1996}).

\bibitem{kabuha2005}
\bibinfo{author}{{Kalberla}, P.~M.~W.} \emph{et~al.}
\newblock \bibinfo{title}{{The Leiden/Argentine/Bonn (LAB) Survey of Galactic
  HI. Final data release of the combined LDS and IAR surveys with improved
  stray-radiation corrections}}.
\newblock \emph{\bibinfo{journal}{\aap}} \textbf{\bibinfo{volume}{440}},
  \bibinfo{pages}{775--782} (\bibinfo{year}{2005}).
\newblock \eprint{arXiv:astro-ph/0504140}.

\bibitem{wialmc2000}
\bibinfo{author}{{Wilms}, J.}, \bibinfo{author}{{Allen}, A.} \&
  \bibinfo{author}{{McCray}, R.}
\newblock \bibinfo{title}{{On the Absorption of X-Rays in the Interstellar
  Medium}}.
\newblock \emph{\bibinfo{journal}{\apj}} \textbf{\bibinfo{volume}{542}},
  \bibinfo{pages}{914--924} (\bibinfo{year}{2000}).
\newblock \eprint{astro-ph/0008425}.


\end{thebibliography}

\clearpage

\begin{thebibliography}{10}
\expandafter\ifx\csname url\endcsname\relax
  \def\url#1{\texttt{#1}}\fi
\expandafter\ifx\csname urlprefix\endcsname\relax\def\urlprefix{URL }\fi
\providecommand{\bibinfo}[2]{#2}
\providecommand{\eprint}[2][]{\url{#2}}
\setcounter{enumiv}{51}

\bibitem{caem2004}
\bibinfo{author}{{Cappellari}, M.} \& \bibinfo{author}{{Emsellem}, E.}
\newblock \bibinfo{title}{{Parametric Recovery of Line-of-Sight Velocity
  Distributions from Absorption-Line Spectra of Galaxies via Penalized
  Likelihood}}.
\newblock \emph{\bibinfo{journal}{\pasp}} \textbf{\bibinfo{volume}{116}},
  \bibinfo{pages}{138--147} (\bibinfo{year}{2004}).
\newblock \eprint{astro-ph/0312201}.

\bibitem{mast2011}
\bibinfo{author}{{Maraston}, C.} \& \bibinfo{author}{{Str{\"o}mb{\"a}ck}, G.}
\newblock \bibinfo{title}{{Stellar population models at high spectral
  resolution}}.
\newblock \emph{\bibinfo{journal}{\mnras}} \textbf{\bibinfo{volume}{418}},
  \bibinfo{pages}{2785--2811} (\bibinfo{year}{2011}).
\newblock \eprint{1109.0543}.

\bibitem{prso2001}
\bibinfo{author}{{Prugniel}, P.} \& \bibinfo{author}{{Soubiran}, C.}
\newblock \bibinfo{title}{{A database of high and medium-resolution stellar
  spectra}}.
\newblock \emph{\bibinfo{journal}{\aap}} \textbf{\bibinfo{volume}{369}},
  \bibinfo{pages}{1048--1057} (\bibinfo{year}{2001}).
\newblock \eprint{astro-ph/0101378}.

\bibitem{mastth2009}
\bibinfo{author}{{Maraston}, C.}, \bibinfo{author}{{Str{\"o}mb{\"a}ck}, G.},
  \bibinfo{author}{{Thomas}, D.}, \bibinfo{author}{{Wake}, D.~A.} \&
  \bibinfo{author}{{Nichol}, R.~C.}
\newblock \bibinfo{title}{{Modelling the colour evolution of luminous red
  galaxies - improvements with empirical stellar spectra}}.
\newblock \emph{\bibinfo{journal}{\mnras}} \textbf{\bibinfo{volume}{394}},
  \bibinfo{pages}{L107--L111} (\bibinfo{year}{2009}).
\newblock \eprint{0809.1867}.

\bibitem{scfida1998}
\bibinfo{author}{{Schlegel}, D.~J.}, \bibinfo{author}{{Finkbeiner}, D.~P.} \&
  \bibinfo{author}{{Davis}, M.}
\newblock \bibinfo{title}{{Maps of Dust Infrared Emission for Use in Estimation
  of Reddening and Cosmic Microwave Background Radiation Foregrounds}}.
\newblock \emph{\bibinfo{journal}{\apj}} \textbf{\bibinfo{volume}{500}},
  \bibinfo{pages}{525--+} (\bibinfo{year}{1998}).
\newblock \eprint{arXiv:astro-ph/9710327}.

\bibitem{baphte1981}
\bibinfo{author}{{Baldwin}, J.~A.}, \bibinfo{author}{{Phillips}, M.~M.} \&
  \bibinfo{author}{{Terlevich}, R.}
\newblock \bibinfo{title}{{Classification parameters for the emission-line
  spectra of extragalactic objects}}.
\newblock \emph{\bibinfo{journal}{\pasp}} \textbf{\bibinfo{volume}{93}},
  \bibinfo{pages}{5--19} (\bibinfo{year}{1981}).

\bibitem{veos1987}
\bibinfo{author}{{Veilleux}, S.} \& \bibinfo{author}{{Osterbrock}, D.~E.}
\newblock \bibinfo{title}{{Spectral classification of emission-line galaxies}}.
\newblock \emph{\bibinfo{journal}{\apjs}} \textbf{\bibinfo{volume}{63}},
  \bibinfo{pages}{295--310} (\bibinfo{year}{1987}).

\bibitem{ke1998}
\bibinfo{author}{{Kennicutt}, R.~C., Jr.}
\newblock \bibinfo{title}{{Star Formation in Galaxies Along the Hubble
  Sequence}}.
\newblock \emph{\bibinfo{journal}{\araa}} \textbf{\bibinfo{volume}{36}},
  \bibinfo{pages}{189--232} (\bibinfo{year}{1998}).
\newblock \eprint{astro-ph/9807187}.

\bibitem{koxuzh2008}
\bibinfo{author}{{Komossa}, S.}, \bibinfo{author}{{Xu}, D.},
  \bibinfo{author}{{Zhou}, H.}, \bibinfo{author}{{Storchi-Bergmann}, T.} \&
  \bibinfo{author}{{Binette}, L.}
\newblock \bibinfo{title}{{On the Nature of Seyfert Galaxies with High [O III]
  {$\lambda$}5007 Blueshifts}}.
\newblock \emph{\bibinfo{journal}{\apj}} \textbf{\bibinfo{volume}{680}},
  \bibinfo{pages}{926--938} (\bibinfo{year}{2008}).
\newblock \eprint{0803.0240}.

\bibitem{mualfi2013}
\bibinfo{author}{{Mullaney}, J.~R.} \emph{et~al.}
\newblock \bibinfo{title}{{Narrow-line region gas kinematics of 24 264
  optically selected AGN: the radio connection}}.
\newblock \emph{\bibinfo{journal}{\mnras}} \textbf{\bibinfo{volume}{433}},
  \bibinfo{pages}{622--638} (\bibinfo{year}{2013}).
\newblock \eprint{1305.0263}.

\bibitem{haalmu2014}
\bibinfo{author}{{Harrison}, C.~M.}, \bibinfo{author}{{Alexander}, D.~M.},
  \bibinfo{author}{{Mullaney}, J.~R.} \& \bibinfo{author}{{Swinbank}, A.~M.}
\newblock \bibinfo{title}{{Kiloparsec-scale outflows are prevalent among
  luminous AGN: outflows and feedback in the context of the overall AGN
  population}}.
\newblock \emph{\bibinfo{journal}{\mnras}} \textbf{\bibinfo{volume}{441}},
  \bibinfo{pages}{3306--3347} (\bibinfo{year}{2014}).
\newblock \eprint{1403.3086}.

\bibitem{ruvesa2005}
\bibinfo{author}{{Rupke}, D.~S.}, \bibinfo{author}{{Veilleux}, S.} \&
  \bibinfo{author}{{Sanders}, D.~B.}
\newblock \bibinfo{title}{{Outflows in Infrared-Luminous Starbursts at z < 0.5.
  II. Analysis and Discussion}}.
\newblock \emph{\bibinfo{journal}{\apjs}} \textbf{\bibinfo{volume}{160}},
  \bibinfo{pages}{115--148} (\bibinfo{year}{2005}).
\newblock \eprint{astro-ph/0506611}.

\bibitem{hokopr2015}
\bibinfo{author}{{Holoien}, T.~W.-S.} \emph{et~al.}
\newblock \bibinfo{title}{{Six Months of Multi-Wavelength Follow-up of the
  Tidal Disruption Candidate ASASSN-14li and Implied TDE Rates from ASAS-SN}}.
\newblock \emph{\bibinfo{journal}{ArXiv e-prints}}  (\bibinfo{year}{2015}).
\newblock \eprint{1507.01598}.

\bibitem{sarees2012}
\bibinfo{author}{{Saxton}, R.~D.} \emph{et~al.}
\newblock \bibinfo{title}{{A tidal disruption-like X-ray flare from the
  quiescent galaxy SDSS J120136.02+300305.5}}.
\newblock \emph{\bibinfo{journal}{\aap}} \textbf{\bibinfo{volume}{541}},
  \bibinfo{pages}{A106} (\bibinfo{year}{2012}).
\newblock \eprint{1202.5900}.

\bibitem{remc2006}
\bibinfo{author}{{Remillard}, R.~A.} \& \bibinfo{author}{{McClintock}, J.~E.}
\newblock \bibinfo{title}{{X-Ray Properties of Black-Hole Binaries}}.
\newblock \emph{\bibinfo{journal}{\araa}} \textbf{\bibinfo{volume}{44}},
  \bibinfo{pages}{49--92} (\bibinfo{year}{2006}).
\newblock \eprint{astro-ph/0606352}.

\bibitem{dogiku2007}
\bibinfo{author}{{Done}, C.}, \bibinfo{author}{{Gierli{\'n}ski}, M.} \&
  \bibinfo{author}{{Kubota}, A.}
\newblock \bibinfo{title}{{Modelling the behaviour of accretion flows in X-ray
  binaries. Everything you always wanted to know about accretion but were
  afraid to ask}}.
\newblock \emph{\bibinfo{journal}{\aapr}} \textbf{\bibinfo{volume}{15}},
  \bibinfo{pages}{1--66} (\bibinfo{year}{2007}).
\newblock \eprint{0708.0148}.

\bibitem{dodaji2012}
\bibinfo{author}{{Done}, C.}, \bibinfo{author}{{Davis}, S.~W.},
  \bibinfo{author}{{Jin}, C.}, \bibinfo{author}{{Blaes}, O.} \&
  \bibinfo{author}{{Ward}, M.}
\newblock \bibinfo{title}{{Intrinsic disc emission and the soft X-ray excess in
  active galactic nuclei}}.
\newblock \emph{\bibinfo{journal}{\mnras}} \textbf{\bibinfo{volume}{420}},
  \bibinfo{pages}{1848--1860} (\bibinfo{year}{2012}).
\newblock \eprint{1107.5429}.

\bibitem{albegu2016}
\bibinfo{author}{{Alexander}, K.~D.}, \bibinfo{author}{{Berger}, E.},
  \bibinfo{author}{{Guillochon}, J.}, \bibinfo{author}{{Zauderer}, B.~A.} \&
  \bibinfo{author}{{Williams}, P.~K.~G.}
\newblock \bibinfo{title}{{Discovery of an Outflow from Radio Observations of
  the Tidal Disruption Event ASASSN-14li}}.
\newblock \emph{\bibinfo{journal}{\apjl}} \textbf{\bibinfo{volume}{819}},
  \bibinfo{pages}{L25} (\bibinfo{year}{2016}).
\newblock \eprint{1510.01226}.

\bibitem{limair2015}
\bibinfo{author}{{Lin}, D.} \emph{et~al.}
\newblock \bibinfo{title}{{An Ultrasoft X-Ray Flare from 3XMM J152130.7+074916:
  A Tidal Disruption Event Candidate}}.
\newblock \emph{\bibinfo{journal}{\apj}} \textbf{\bibinfo{volume}{811}},
  \bibinfo{pages}{43} (\bibinfo{year}{2015}).
\newblock \eprint{1509.00840}.

\bibitem{gura2013}
\bibinfo{author}{{Guillochon}, J.} \& \bibinfo{author}{{Ramirez-Ruiz}, E.}
\newblock \bibinfo{title}{{Hydrodynamical Simulations to Determine the Feeding
  Rate of Black Holes by the Tidal Disruption of Stars: The Importance of the
  Impact Parameter and Stellar Structure}}.
\newblock \emph{\bibinfo{journal}{\apj}} \textbf{\bibinfo{volume}{767}},
  \bibinfo{pages}{25} (\bibinfo{year}{2013}).
\newblock \eprint{1206.2350}.

\bibitem{lireho2009}
\bibinfo{author}{{Lin}, D.}, \bibinfo{author}{{Remillard}, R.~A.} \&
  \bibinfo{author}{{Homan}, J.}
\newblock \bibinfo{title}{{Spectral States of XTE J1701-462: Link Between Z and
  Atoll Sources}}.
\newblock \emph{\bibinfo{journal}{\apj}} \textbf{\bibinfo{volume}{696}},
  \bibinfo{pages}{1257--1277} (\bibinfo{year}{2009}).
\newblock \eprint{0901.0031}.

\bibitem{magura2012}
\bibinfo{author}{{MacLeod}, M.}, \bibinfo{author}{{Guillochon}, J.} \&
  \bibinfo{author}{{Ramirez-Ruiz}, E.}
\newblock \bibinfo{title}{{The Tidal Disruption of Giant Stars and their
  Contribution to the Flaring Supermassive Black Hole Population}}.
\newblock \emph{\bibinfo{journal}{\apj}} \textbf{\bibinfo{volume}{757}},
  \bibinfo{pages}{134} (\bibinfo{year}{2012}).
\newblock \eprint{1206.2922}.

\bibitem{maragr2013}
\bibinfo{author}{{MacLeod}, M.}, \bibinfo{author}{{Ramirez-Ruiz}, E.},
  \bibinfo{author}{{Grady}, S.} \& \bibinfo{author}{{Guillochon}, J.}
\newblock \bibinfo{title}{{Spoon-feeding Giant Stars to Supermassive Black
  Holes: Episodic Mass Transfer from Evolving Stars and their Contribution to
  the Quiescent Activity of Galactic Nuclei}}.
\newblock \emph{\bibinfo{journal}{\apj}} \textbf{\bibinfo{volume}{777}},
  \bibinfo{pages}{133} (\bibinfo{year}{2013}).
\newblock \eprint{1307.2900}.

\bibitem{voasbo1999}
\bibinfo{author}{{Voges}, W.} \emph{et~al.}
\newblock \bibinfo{title}{{The ROSAT all-sky survey bright source catalogue}}.
\newblock \emph{\bibinfo{journal}{\aap}} \textbf{\bibinfo{volume}{349}},
  \bibinfo{pages}{389--405} (\bibinfo{year}{1999}).
\newblock \eprint{arXiv:astro-ph/9909315}.

\bibitem{hokopr2016}
\bibinfo{author}{{Holoien}, T.~W.-S.} \emph{et~al.}
\newblock \bibinfo{title}{{ASASSN-15oi: A Rapidly Evolving, Luminous Tidal
  Disruption Event at 216 Mpc}}.
\newblock \emph{\bibinfo{journal}{ArXiv e-prints}}  (\bibinfo{year}{2016}).
\newblock \eprint{1602.01088}.

\bibitem{labima2009}
\bibinfo{author}{{Lamastra}, A.} \emph{et~al.}
\newblock \bibinfo{title}{{The bolometric luminosity of type 2 AGN from
  extinction-corrected [OIII]. No evidence of Eddington-limited sources}}.
\newblock \emph{\bibinfo{journal}{\aap}} \textbf{\bibinfo{volume}{504}},
  \bibinfo{pages}{73--79} (\bibinfo{year}{2009}).
\newblock \eprint{0905.4439}.

\bibitem{liweba2012}
\bibinfo{author}{{Lin}, D.}, \bibinfo{author}{{Webb}, N.~A.} \&
  \bibinfo{author}{{Barret}, D.}
\newblock \bibinfo{title}{{Classification of X-Ray Sources in the XMM-Newton
  Serendipitous Source Catalog}}.
\newblock \emph{\bibinfo{journal}{\apj}} \textbf{\bibinfo{volume}{756}},
  \bibinfo{pages}{27} (\bibinfo{year}{2012}).
\newblock \eprint{1207.1913}.

\bibitem{grkosc2013}
\bibinfo{author}{{Grupe}, D.} \emph{et~al.}
\newblock \bibinfo{title}{{Strong UV and X-Ray Variability of the Narrow Line
  Seyfert 1 Galaxy WPVS 007: on the Nature of the X-Ray Low State}}.
\newblock \emph{\bibinfo{journal}{\aj}} \textbf{\bibinfo{volume}{146}},
  \bibinfo{pages}{78} (\bibinfo{year}{2013}).
\newblock \eprint{1307.7713}.

\bibitem{tekaaw2012}
\bibinfo{author}{{Terashima}, Y.}, \bibinfo{author}{{Kamizasa}, N.},
  \bibinfo{author}{{Awaki}, H.}, \bibinfo{author}{{Kubota}, A.} \&
  \bibinfo{author}{{Ueda}, Y.}
\newblock \bibinfo{title}{{A Candidate Active Galactic Nucleus with a Pure Soft
  Thermal X-Ray Spectrum}}.
\newblock \emph{\bibinfo{journal}{\apj}} \textbf{\bibinfo{volume}{752}},
  \bibinfo{pages}{154} (\bibinfo{year}{2012}).
\newblock \eprint{1205.2774}.

\bibitem{hokite2012}
\bibinfo{author}{{Ho}, L.~C.}, \bibinfo{author}{{Kim}, M.} \&
  \bibinfo{author}{{Terashima}, Y.}
\newblock \bibinfo{title}{{The Low-mass, Highly Accreting Black Hole Associated
  with the Active Galactic Nucleus 2XMM J123103.2+110648}}.
\newblock \emph{\bibinfo{journal}{\apjl}} \textbf{\bibinfo{volume}{759}},
  \bibinfo{pages}{L16} (\bibinfo{year}{2012}).
\newblock \eprint{1210.0440}.

\bibitem{liirgo2013}
\bibinfo{author}{{Lin}, D.}, \bibinfo{author}{{Irwin}, J.~A.},
  \bibinfo{author}{{Godet}, O.}, \bibinfo{author}{{Webb}, N.~A.} \&
  \bibinfo{author}{{Barret}, D.}
\newblock \bibinfo{title}{{A \~{} 3.8 hr Periodicity from an Ultrasoft Active
  Galactic Nucleus Candidate}}.
\newblock \emph{\bibinfo{journal}{\apjl}} \textbf{\bibinfo{volume}{776}},
  \bibinfo{pages}{L10} (\bibinfo{year}{2013}).
\newblock \eprint{1309.4440}.

\bibitem{liweba2014}
\bibinfo{author}{{Lin}, D.}, \bibinfo{author}{{Webb}, N.~A.} \&
  \bibinfo{author}{{Barret}, D.}
\newblock \bibinfo{title}{{Classification of X-Ray Sources in the XMM-Newton
  Serendipitous Source Catalog: Objects of Special Interest}}.
\newblock \emph{\bibinfo{journal}{\apj}} \textbf{\bibinfo{volume}{780}},
  \bibinfo{pages}{39} (\bibinfo{year}{2014}).
\newblock \eprint{1309.0509}.

\bibitem{misaro2013}
\bibinfo{author}{{Miniutti}, G.} \emph{et~al.}
\newblock \bibinfo{title}{{A high Eddington-ratio, true Seyfert 2 galaxy
  candidate: implications for broad-line region models}}.
\newblock \emph{\bibinfo{journal}{\mnras}} \textbf{\bibinfo{volume}{433}},
  \bibinfo{pages}{1764--1777} (\bibinfo{year}{2013}).
\newblock \eprint{1305.3284}.

\bibitem{samoko2015}
\bibinfo{author}{{Saxton}, R.~D.}, \bibinfo{author}{{Motta}, S.~E.},
  \bibinfo{author}{{Komossa}, S.} \& \bibinfo{author}{{Read}, A.~M.}
\newblock \bibinfo{title}{{Was the soft X-ray flare in NGC 3599 due to an AGN
  disc instability or a delayed tidal disruption event?}}
\newblock \emph{\bibinfo{journal}{\mnras}} \textbf{\bibinfo{volume}{454}},
  \bibinfo{pages}{2798--2803} (\bibinfo{year}{2015}).
\newblock \eprint{1509.05193}.

\bibitem{grkosa2015}
\bibinfo{author}{{Grupe}, D.}, \bibinfo{author}{{Komossa}, S.} \&
  \bibinfo{author}{{Saxton}, R.}
\newblock \bibinfo{title}{{IC 3599 Did It Again: A Second Outburst of the X-Ray
  Transient Seyfert 1.9 Galaxy}}.
\newblock \emph{\bibinfo{journal}{\apjl}} \textbf{\bibinfo{volume}{803}},
  \bibinfo{pages}{L28} (\bibinfo{year}{2015}).
\newblock \eprint{1504.01389}.

\bibitem{camaco2015}
\bibinfo{author}{{Campana}, S.} \emph{et~al.}
\newblock \bibinfo{title}{{Multiple tidal disruption flares in the active
  galaxy IC 3599}}.
\newblock \emph{\bibinfo{journal}{\aap}} \textbf{\bibinfo{volume}{581}},
  \bibinfo{pages}{A17} (\bibinfo{year}{2015}).
\newblock \eprint{1502.07184}.

\bibitem{shprgr2014}
\bibinfo{author}{{Shappee}, B.~J.} \emph{et~al.}
\newblock \bibinfo{title}{{The Man behind the Curtain: X-Rays Drive the UV
  through NIR Variability in the 2013 Active Galactic Nucleus Outburst in NGC
  2617}}.
\newblock \emph{\bibinfo{journal}{\apj}} \textbf{\bibinfo{volume}{788}},
  \bibinfo{pages}{48} (\bibinfo{year}{2014}).
\newblock \eprint{1310.2241}.

\bibitem{lacamo2015}
\bibinfo{author}{{LaMassa}, S.~M.} \emph{et~al.}
\newblock \bibinfo{title}{{The Discovery of the First Changing Look Quasar: New
  Insights Into the Physics and Phenomenology of Active Galactic Nucleus}}.
\newblock \emph{\bibinfo{journal}{\apj}} \textbf{\bibinfo{volume}{800}},
  \bibinfo{pages}{144} (\bibinfo{year}{2015}).
\newblock \eprint{1412.2136}.

\bibitem{pakoko2016}
\bibinfo{author}{{Parker}, M.~L.} \emph{et~al.}
\newblock \bibinfo{title}{{The detection and X-ray view of the changing look
  AGN HE 1136-2304}}.
\newblock \emph{\bibinfo{journal}{\mnras}} \textbf{\bibinfo{volume}{461}},
  \bibinfo{pages}{1927--1936} (\bibinfo{year}{2016}).
\newblock \eprint{1606.04955}.

\bibitem{risael2009}
\bibinfo{author}{{Risaliti}, G.} \emph{et~al.}
\newblock \bibinfo{title}{{The XMM-Newton long look of NGC 1365: uncovering of
  the obscured X-ray source}}.
\newblock \emph{\bibinfo{journal}{\mnras}} \textbf{\bibinfo{volume}{393}},
  \bibinfo{pages}{L1--L5} (\bibinfo{year}{2009}).
\newblock \eprint{0811.1594}.

\bibitem{rimiel2009}
\bibinfo{author}{{Risaliti}, G.} \emph{et~al.}
\newblock \bibinfo{title}{{Variable Partial Covering and A Relativistic Iron
  Line in NGC 1365}}.
\newblock \emph{\bibinfo{journal}{\apj}} \textbf{\bibinfo{volume}{696}},
  \bibinfo{pages}{160--171} (\bibinfo{year}{2009}).
\newblock \eprint{0901.4809}.

\bibitem{fisa1989}
\bibinfo{author}{{Filippenko}, A.~V.} \& \bibinfo{author}{{Sargent}, W.~L.~W.}
\newblock \bibinfo{title}{{Discovery of an extremely low luminosity Seyfert 1
  nucleus in the dwarf galaxy NGC 4395}}.
\newblock \emph{\bibinfo{journal}{\apjl}} \textbf{\bibinfo{volume}{342}},
  \bibinfo{pages}{L11--L14} (\bibinfo{year}{1989}).

\bibitem{bahoru2004}
\bibinfo{author}{{Barth}, A.~J.}, \bibinfo{author}{{Ho}, L.~C.},
  \bibinfo{author}{{Rutledge}, R.~E.} \& \bibinfo{author}{{Sargent}, W.~L.~W.}
\newblock \bibinfo{title}{{POX 52: A Dwarf Seyfert 1 Galaxy with an
  Intermediate-Mass Black Hole}}.
\newblock \emph{\bibinfo{journal}{\apj}} \textbf{\bibinfo{volume}{607}},
  \bibinfo{pages}{90--102} (\bibinfo{year}{2004}).
\newblock \eprint{astro-ph/0402110}.

\bibitem{resijo2011}
\bibinfo{author}{{Reines}, A.~E.}, \bibinfo{author}{{Sivakoff}, G.~R.},
  \bibinfo{author}{{Johnson}, K.~E.} \& \bibinfo{author}{{Brogan}, C.~L.}
\newblock \bibinfo{title}{{An actively accreting massive black hole in the
  dwarf starburst galaxy Henize2-10}}.
\newblock \emph{\bibinfo{journal}{\nat}} \textbf{\bibinfo{volume}{470}},
  \bibinfo{pages}{66--68} (\bibinfo{year}{2011}).
\newblock \eprint{1101.1309}.

\bibitem{pagogr2016}
\bibinfo{author}{{Pardo}, K.} \emph{et~al.}
\newblock \bibinfo{title}{{X-ray Detected Active Galactic Nuclei in Dwarf
  Galaxies at $0<1$}}.
\newblock \emph{\bibinfo{journal}{ArXiv e-prints}}  (\bibinfo{year}{2016}).
\newblock \eprint{1603.01622}.

\bibitem{fa1999}
\bibinfo{author}{{Fabian}, A.~C.}
\newblock \bibinfo{title}{{The obscured growth of massive black holes}}.
\newblock \emph{\bibinfo{journal}{\mnras}} \textbf{\bibinfo{volume}{308}},
  \bibinfo{pages}{L39--L43} (\bibinfo{year}{1999}).
\newblock \eprint{astro-ph/9908064}.

\bibitem{sire1998}
\bibinfo{author}{{Silk}, J.} \& \bibinfo{author}{{Rees}, M.~J.}
\newblock \bibinfo{title}{{Quasars and galaxy formation}}.
\newblock \emph{\bibinfo{journal}{\aap}} \textbf{\bibinfo{volume}{331}},
  \bibinfo{pages}{L1--L4} (\bibinfo{year}{1998}).
\newblock \eprint{astro-ph/9801013}.

\bibitem{ko2012}
\bibinfo{author}{{Komossa}, S.}
\newblock \bibinfo{title}{{Tidal disruption of stars by supermassive black
  holes: The X-ray view}}.
\newblock In \emph{\bibinfo{booktitle}{European Physical Journal Web of
  Conferences}}, vol.~\bibinfo{volume}{39} of \emph{\bibinfo{series}{European
  Physical Journal Web of Conferences}}, \bibinfo{pages}{2001}
  (\bibinfo{year}{2012}).

\bibitem{ko2015}
\bibinfo{author}{{Komossa}, S.}
\newblock \bibinfo{title}{{Tidal disruption of stars by supermassive black
  holes: Status of observations}}.
\newblock \emph{\bibinfo{journal}{JHEA}} \textbf{\bibinfo{volume}{7}},
  \bibinfo{pages}{148--157} (\bibinfo{year}{2015}).
\newblock \eprint{1505.01093}.

\bibitem{grthle1999}
\bibinfo{author}{{Grupe}, D.}, \bibinfo{author}{{Thomas}, H.} \&
  \bibinfo{author}{{Leighly}, K.~M.}
\newblock \bibinfo{title}{{RX J1624.9+7554: a new X-ray transient AGN}}.
\newblock \emph{\bibinfo{journal}{\aap}} \textbf{\bibinfo{volume}{350}},
  \bibinfo{pages}{L31--L34} (\bibinfo{year}{1999}).
\newblock \eprint{arXiv:astro-ph/9909101}.

\bibitem{kogr1999}
\bibinfo{author}{{Komossa}, S.} \& \bibinfo{author}{{Greiner}, J.}
\newblock \bibinfo{title}{{Discovery of a giant and luminous X-ray outburst
  from the optically inactive galaxy pair RX J1242.6-1119}}.
\newblock \emph{\bibinfo{journal}{\aap}} \textbf{\bibinfo{volume}{349}},
  \bibinfo{pages}{L45--L48} (\bibinfo{year}{1999}).
\newblock \eprint{arXiv:astro-ph/9908216}.

\bibitem{koba1999}
\bibinfo{author}{{Komossa}, S.} \& \bibinfo{author}{{Bade}, N.}
\newblock \bibinfo{title}{{The giant X-ray outbursts in NGC 5905 and IC 3599:()
  hfill Follow-up observations and outburst scenarios}}.
\newblock \emph{\bibinfo{journal}{\aap}} \textbf{\bibinfo{volume}{343}},
  \bibinfo{pages}{775--787} (\bibinfo{year}{1999}).
\newblock \eprint{arXiv:astro-ph/9901141}.

\bibitem{essako2008}
\bibinfo{author}{{Esquej}, P.} \emph{et~al.}
\newblock \bibinfo{title}{{Evolution of tidal disruption candidates discovered
  by XMM-Newton}}.
\newblock \emph{\bibinfo{journal}{\aap}} \textbf{\bibinfo{volume}{489}},
  \bibinfo{pages}{543--554} (\bibinfo{year}{2008}).
\newblock \eprint{0807.4452}.

\bibitem{licagr2011}
\bibinfo{author}{{Lin}, D.} \emph{et~al.}
\newblock \bibinfo{title}{{Discovery of an Ultrasoft X-Ray Transient Source in
  the 2XMM Catalog: A Tidal Disruption Event Candidate}}.
\newblock \emph{\bibinfo{journal}{\apj}} \textbf{\bibinfo{volume}{738}},
  \bibinfo{pages}{52--+} (\bibinfo{year}{2011}).
\newblock \eprint{1106.0744}.

\bibitem{mauler2010}
\bibinfo{author}{{Maksym}, W.~P.}, \bibinfo{author}{{Ulmer}, M.~P.} \&
  \bibinfo{author}{{Eracleous}, M.}
\newblock \bibinfo{title}{{A Tidal Disruption Flare in A1689 from an Archival
  X-ray Survey of Galaxy Clusters}}.
\newblock \emph{\bibinfo{journal}{\apj}} \textbf{\bibinfo{volume}{722}},
  \bibinfo{pages}{1035--1050} (\bibinfo{year}{2010}).
\newblock \eprint{1008.4140}.

\bibitem{maliir2014}
\bibinfo{author}{{Maksym}, W.~P.}, \bibinfo{author}{{Lin}, D.} \&
  \bibinfo{author}{{Irwin}, J.~A.}
\newblock \bibinfo{title}{{RBS 1032: A Tidal Disruption Event in Another Dwarf
  Galaxy?}}
\newblock \emph{\bibinfo{journal}{\apjl}} \textbf{\bibinfo{volume}{792}},
  \bibinfo{pages}{L29} (\bibinfo{year}{2014}).
\newblock \eprint{1407.2928}.

\bibitem{doceco2014}
\bibinfo{author}{{Donato}, D.} \emph{et~al.}
\newblock \bibinfo{title}{{A Tidal Disruption Event in a nearby Galaxy Hosting
  an Intermediate Mass Black Hole}}.
\newblock \emph{\bibinfo{journal}{\apj}} \textbf{\bibinfo{volume}{781}},
  \bibinfo{pages}{59} (\bibinfo{year}{2014}).
\newblock \eprint{1311.6162}.

\bibitem{blgime2011}
\bibinfo{author}{{Bloom}, J.~S.} \emph{et~al.}
\newblock \bibinfo{title}{{A Possible Relativistic Jetted Outburst from a
  Massive Black Hole Fed by a Tidally Disrupted Star}}.
\newblock \emph{\bibinfo{journal}{Science}} \textbf{\bibinfo{volume}{333}},
  \bibinfo{pages}{203--} (\bibinfo{year}{2011}).
\newblock \eprint{1104.3257}.

\bibitem{bukegh2011}
\bibinfo{author}{{Burrows}, D.~N.} \emph{et~al.}
\newblock \bibinfo{title}{{Relativistic jet activity from the tidal disruption
  of a star by a massive black hole}}.
\newblock \emph{\bibinfo{journal}{\nat}} \textbf{\bibinfo{volume}{476}},
  \bibinfo{pages}{421--424} (\bibinfo{year}{2011}).
\newblock \eprint{1104.4787}.

\bibitem{zabeso2011}
\bibinfo{author}{{Zauderer}, B.~A.} \emph{et~al.}
\newblock \bibinfo{title}{{Birth of a relativistic outflow in the unusual
  {$\gamma$}-ray transient Swift J164449.3+573451}}.
\newblock \emph{\bibinfo{journal}{\nat}} \textbf{\bibinfo{volume}{476}},
  \bibinfo{pages}{425--428} (\bibinfo{year}{2011}).
\newblock \eprint{1106.3568}.

\bibitem{cekrho2012}
\bibinfo{author}{{Cenko}, S.~B.} \emph{et~al.}
\newblock \bibinfo{title}{{Swift J2058.4+0516: Discovery of a Possible Second
  Relativistic Tidal Disruption Flare?}}
\newblock \emph{\bibinfo{journal}{\apj}} \textbf{\bibinfo{volume}{753}},
  \bibinfo{pages}{77} (\bibinfo{year}{2012}).
\newblock \eprint{1107.5307}.

\bibitem{brlest2015}
\bibinfo{author}{{Brown}, G.~C.} \emph{et~al.}
\newblock \bibinfo{title}{{Swift J1112.2-8238: a candidate relativistic tidal
  disruption flare}}.
\newblock \emph{\bibinfo{journal}{\mnras}} \textbf{\bibinfo{volume}{452}},
  \bibinfo{pages}{4297--4306} (\bibinfo{year}{2015}).
\newblock \eprint{1507.03582}.

\bibitem{kegrka2006}
\bibinfo{author}{{Kewley}, L.~J.}, \bibinfo{author}{{Groves}, B.},
  \bibinfo{author}{{Kauffmann}, G.} \& \bibinfo{author}{{Heckman}, T.}
\newblock \bibinfo{title}{{The host galaxies and classification of active
  galactic nuclei}}.
\newblock \emph{\bibinfo{journal}{\mnras}} \textbf{\bibinfo{volume}{372}},
  \bibinfo{pages}{961--976} (\bibinfo{year}{2006}).
\newblock \eprint{astro-ph/0605681}.
\end{thebibliography}
\end{document}